\documentclass[10pt,article,showpacs,showkeys,amsmath,amssymb,
 aps,prc,twocolumn]{revtex4}

\usepackage{graphicx}
\usepackage{dcolumn}
\usepackage{bm}
\usepackage{xcolor}
\usepackage{lineno}
\begin{document}

\preprint{PRC/Relativistic Fluid Dynamics}

\title{Third order viscous hydrodynamics from the entropy four current}

\author{Mohammed Younus}
\email{younus.presi@gmail.com}
\affiliation{Department of Physics, Nelson Mandela University, Port Elizabeth, 6031, South Africa}%
\author{Azwinndini Muronga}%
\email{Azwinndini.Muronga@mandela.ac.za}
\affiliation{Faculty of Science, Nelson Mandela University, Port Elizabeth, 6031, South Africa}%


\date{\today}

\begin{abstract}
Non-equilibrium dynamics for relativistic fluid or quark gluon plasma (QGP) have already been calculated earlier upto third order using both kinetic and thermodynamic approaches. Calculations presented in this manuscript are based on thermodynamics principles. The expressions for third order dissipative fluxes have been derived from equation for entropy 4-current developed earlier by A. Muronga. The relaxation equations in the present work have been developed in a simple Bjorken (1+1)D scenario and Eckart frame. The relaxation equations are found to have slightly different values for the coupling coefficients as compared to calculations from earlier models. The solutions to the differential equations have been found to be sensitive to values of these coefficients. The shear relaxation equations derived in third order theory are discussed term by term. Effects of third order theory on shear relaxation time has been discussed. Thermodynamic quantities related to hot and dense matter have been calculated as functions of proper time. Moreover, various initial conditions for the relaxation equations have been assumed to study their effects on above mentioned observables. A LHC QGP formation time of $\displaystyle{\tau_0}$ = 0.4 fm/c and temperature of T$_0$ = 500 MeV have been assumed.
\end{abstract}

\keywords{Relativistic hydrodynamics, quark gluon plasma, third order viscous hydrodynamics}
\maketitle


\section{Introduction}

High energy heavy ion collisions offer the opportunity to study the properties of hot and dense quark gluon plasma (QGP)~\cite{Harris:1996zx}. At LHC-CERN, experimental results have suggested the formation of a relativistic fluid with observables on particle production and have thus confirmed formation of QGP~\cite{Adler:2003kt,Li:2010ew,Abelev:2014pua}. In order to study the system of relativistic fluid evolving through space and time, one must use transport equations. Transport equations don't merely transport particle distributions without any medium effects or particle interaction but also include various physical and non-equilibrium processes such as dissipations, collisions and radiations. The non-equilibrium phenomena are particularly interesting because of the various transport coefficients and their relaxation times and length scales which may help us track equilibration of the system. Thus the use of fluid dynamics as one of the approaches in modelling the dynamic evolution of nuclear collisions has been successful in describing many of the observables~\cite{Stoecker:1986ci,Bass:1998vz,Bleicher:1999xi}. However the assumptions and the approximations of the fluid dynamical models have been the source of major uncertainties in predicting the observables.

Many works have been done earlier on relativistic fluid dynamics~\cite{Kapusta:1982va} both from kinetic theory and thermodynamics approaches. The first order theories of relativistic fluid dynamics are due to Eckart et al~\cite{Eckart:1940te}. and to Landau and Lifshitz~\cite{Landau:1959fm}, and had assumptions that the entropy four-current contains only linear terms in dissipative quantities. Consequently we have Fourier-Navier-Stokes equations which might lead to non-causality and propagate viscous and thermal signals with speed greater than that of light. These theories have been extended to include second order equations to meet the causality conditions and have been done at the earliest by Muller and Israel and Stewart. This is also known as second order dissipative theories or Muller-Israel-Stewart theories (M-IS)~\cite{Muller:1967zza,Israel:1976tn,Israel:1979wp}. Recent works to include second order corrections have been done by A. Muronga et al~\cite{Muronga:2001zk,Muronga:2003ta,Muronga:2006zw,Muronga:2006zx}, A. El et. al.~\cite{El:2008yy,El:2010mt} using thermodynamics approaches (entropy principle), while G. S. Denicol et al.~\cite{Denicol:2010xn}, A. Jaiswal et al.~\cite{Jaiswal:2013npa,Jaiswal:2015mxa} used an iterative approach to kinetic Boltzmann equation (BE) to solve the dissipative equations. The results from these various approaches are complimentary~\cite{Bhalerao:2013pza}. However their differences are considerable and depend on the approaches or techniques involved and values of the coefficients in the differential equations. We will return to this issue in the discussion section. Both second and third order hydrodynamics are of contemporary interests as some recent works have called upon these theories in including mass effects and fluid-gravity duality. The predictions showed an increased importance of application of hydrodynamics to both massless and massive fluids as well as to compact systems~\cite{Attems:2017ezz,Jaiswal:2016hex,Saida:2017ojb,Bemfica:2019cop,Ryblewski:2014yxa}.

Other recent works have directed hydrodynamics to the study of attractors which might indicate convergence of hydrodynamics coefficients. The works in references~\cite{Heller:2013fn,Grozdanov:2015kqa} correspond to Ads-CFT, fluid gravity duality whre hydrodynamics attractors upto order, $\displaystyle{n}$ = 240 have been calculated. However the coefficients show a convergence for $\displaystyle{n\,\leq}$ 5, where both second and third order theories are restricted to. The works gave rise application of to Borel resummation techniques to test the convergence. However the works are restricted to massless limit of conformal theories and extending them to massive particles regime could be carried out. Hydrodynamics could be studied at any temperature, or particle density regimes as well as for both massive and massless regimes. Similar extensive works by R. Baier~\cite{Baier:2007ix} with Ads-CFT and Amaresh Jaiswal~\cite{Jaiswal:2019cju} with third order hydrodynamics in BTE-RTA have shown the calculation of attractors and convergence of hydrodynamics coefficients. Indeed it would be interesting to find the maximum order we can reach with hydrodynamics, before it is no more valid.

In the current work we have extended the work done by A. Muronga to third order equations for the dissipative fluids~\cite{Muronga:2010zz} and presented it here as a test model for further development. The calculations are shown briefly in section III. We have compared our calculations and results with earlier calculations by A. El et al. and A. Jaiswal. The results and discussions are reported in section. IV of the current manuscript, followed by conclusions.

\section{EQUILIBRIUM AND NON-EQUILIBRIUM DYNAMICS}
In this section we discuss briefly the equilibrium and non-equilibrium dynamics. The dissipative fluxes which appear as time-dependant variables in the conservation equations, serve as major factors those push the fluid system out of equilibrium state. We also know from the laws of thermodynamics that entropy is conserved in ideal fluid. Because of the absence of dissipative forces, changes in the system are reversible. It is also known from earlier works that despite non-physical nature of perfect fluid, its dynamics are able to explain various phenomena in heavy ion collisions ~\cite{Ollitrault:2008zz}. In real fluids however, dissipative fluxes due to friction, stress and heat flow, cause the system to undergo irreversible processes and lead to increase in entropy. At present relativistic heavy ion collision experiments which are being conducted at LHC-CERN give us the context and opportunity to apply relativistic fluid dynamics and study the dissipative properties of quark gluon plasma (QGP) as well as hadron gases formed from QGP.  Results from AdS-CFT have also suggested a small presence of viscous forces in QGP and give us a lower bound (KSS bound) on $\displaystyle{\frac{\eta}{s}\,=\,\frac{1}{4\pi}}$~\cite{Sinha:2009ev}, where $\displaystyle{\eta}$ is the shear viscosity and $\displaystyle{s}$ is the entropy density of fluid system. 
Developing master equations and their solutions in order to simulate hydrodynamics evolution for relativistic fluids have been a major challenge to researchers. Because of the presence of non-linear terms, formulation of equations becomes non-trivial even in the case of ideal hydrodynamics. Restrictions are also put on the equations by the laws of thermodaynamics, and from the generalized expression from the equations one can derive the expression of equation of state which could describe the thermodynamic states of the fluids. The theoretical formulation for the ideal/non-dissipative hydrodynamics was given by J. D. Bjorken~\cite{Bjorken:1982qr} which gave us the conservation equations for the energy-momentum tensor, number and entropy densities. The Bjorken scaling solution have been applied to first, second and third order theories. The second order theories or M-IS theories use Grad's 14-moment method~\cite{Grad:1949mm,Hiscock:1983zz} and upto second moment of Boltzmann equation as approximation. M-IS included only linear terms in the dissipative equations and neglected non-linear terms as wellas terms such as derivatives of thermodynamic variables. The theories haveve limitations such as reheating of the system etc. Earlier discussions on IS theory in a simple one-dimensional Bjorken scaling expansion have also pointed out some unphysical results such as negative longitudunal pressure at small initial time. Thus higher order corrections along with inclusion of non-linear terms and derivatives of thermodynamic variables are studied to see if these effects can be reduced~\cite{Muronga:2010zz}. To include third order, both thermodynamic approach from Grad's 14-moment approximation by A. El et al.~\cite{El:2009vj} and kinetic approach by A. Jaiswal~\cite{Jaiswal:2013vta,Jaiswal:2014raa} have been done. In both approaches by A. El and A. Jaiswal respectively, a relativistic third order evolution equation for shear stress has been developed to study dissipative dynamics. The results could agree with exact solutions from Boltzmann transport equations. The approach to develop third order theory by A. Muronga takes expansion of Grad's 14-moments upto quadratic terms, and gives full expression for dissipative fluxes and thus entropy 4-current upto third order. We will briefly discuss the formalism in the following section. As mentioned in early seminal works of~\cite{Heller:2013fn,Grozdanov:2015kqa,Romatschke:2017ejr}, that gradient expansion techniques bring in a number of non-linear terms and non-hydrodynamic modes. As to the date these modes are yet to be fully understood. But they are important as they make equations of motion causal. As demonstrated in the above references, particularly through Maxwell-Cattaneo (MC) equations, the presence of non-hydrodynamic modes lead to finite signal propagation speed in contrast to first order or Navier-Stokes (NS) theories. In fact MC theory has derivatives in both space and time. This makes the resulting MC equations of motion hyperbolic, whereas the NS equations are parabolic in nature. Some of the consequences of non-linear modes particularly on relaxation time has been discussed in the latter sections.

\section{FORMALISM}

The basic formulation of relativistic hydrodynamics can be found in the literatures mentioned in previous sections. Here a simple fluid with massless particles and no electromagnetic fields has been considered. The fluid is characterized by

\begin{eqnarray}
&N^\mu(x),\,~~~~ \text{particle 4-current},\\
\nonumber\\
&T^{\mu\nu}(x),\,~~~ \text{energy-momentum tensor},\\
\nonumber\\
&S^\mu(x),\,~~~~ \text{entropy 4-current},
\label{currenttensor1}
\end{eqnarray}

The equations for the conservation of net charge and energy-momentum are given by

\begin{equation}
\partial_\mu N^\mu = 0.
\label{numbercon}
\end{equation}

\begin{equation}
\partial_\mu T^{\mu\nu} = 0.
\label{energycon}
\end{equation}

Also, the second law of thermodynamics dictates,
\begin{equation}
\partial_\mu S^\mu \geqslant 0
\label{entropycon}
\end{equation}

The energy flow vector is similarly defined as
\begin{eqnarray}
W^\mu &=& u_\nu T^{\nu\lambda}\Delta^\mu_\lambda\nonumber\\
      &=& q^\mu + hv^\mu
\end{eqnarray}
where $\displaystyle{h=\frac{\varepsilon+P}{n}}$ is the enthalpy, $\displaystyle{v^\mu}$ is the particle flow vector and $\displaystyle{q^\mu}$ is the heat flow or heat 4-current. In Eckart frame the particle flux $\displaystyle{v^\mu\,=\,0}$ which implies $\displaystyle{W^\mu\,=\,q^\mu}$. In the Landau and Lifshitz frame or energy frame, we have $\displaystyle{W^\mu}\,=\,0$ which implies $\displaystyle{q^\mu\,=\,-hv^\mu}$. 

The net charge 4-current might be written of the form,
\begin{equation}
N^\mu = nu^\mu + v^\mu
\label{number}
\end{equation}
In Eckart frame where there is no particle flux, the particle number density in fluid rest frame is given by $\displaystyle{n=\sqrt{N^\mu N_\mu}}$. It can be shown that $\displaystyle{u^\mu=\frac{N^\mu}{\sqrt{N^\mu N_\mu}}}$ is the fluid 4-velocity such that $\displaystyle{u^\mu u_\mu = 1}$, while the energy momentum tensor can be written as,
\begin{equation}
T^{\mu\nu}=\varepsilon u^\mu u^\nu -(P+\Pi)\Delta^{\mu\nu}+2q^{(\mu}u^{\nu)}+\pi^{\langle\mu\nu\rangle}\,,
\label{energymomentumtensor}
\end{equation}
where $\displaystyle{\varepsilon = u_\mu u_\nu T^{\mu\nu}}$ is the energy density, $P$ is the pressure in fluid rest frame, $\displaystyle{\Pi}$ is the bulk viscous pressure and $\displaystyle{\pi^{\langle\mu\nu\rangle}}$ is shear stress tensor. $\displaystyle{\Delta^{\mu\nu}\,=\,g^{\mu\nu}-u^\mu u^\nu}$ is the projection tensor in 3 D space.

In the present calculation Grad's 14-moments approximation has been used to develop the equations for dissipative fluxes.
A system of relativistic fluid has been considered that departs slightly from the local thermal distribution. The distribution for particles in that system can then be written as
\begin{equation}
f(x,p)=f^{eq}(x,p)[1+\Delta^{eq}\phi (x,p)]\,,
\label{distribution}
\end{equation}
where
\begin{equation}
f^{eq}(x,p)=A_0\frac{1}{e^{\beta_\nu p^\nu-\alpha}-a}\,,
\label{eqdistribution}
\end{equation}
is the equilibrium distribution function. The factor, $\displaystyle{\Delta^{eq}}$ is expressed as $\displaystyle{1+aA_0^{-1}f(x,p)}$ and $\displaystyle{\phi(x,p)}$ is the deviation/departure function. The definition of $\displaystyle{f(x,p)}$ has been used in Eq.~\ref{distribution} for the following expressions for number, energy-momentum tensor and fluxes equations,
\begin{eqnarray}
N^\mu(x)&=&\int{f(x,p)p^\mu\,dw}\nonumber\\
\nonumber\\
T^{\mu\nu}(x)&=&\int{f(x,p)p^\mu p^\nu\,dw},\,\text{and}\nonumber\\
\nonumber\\
F^{\mu\nu\lambda}(x)&=&\int{f(x,p)p^\mu p^\nu p^\lambda\,dw}\,,
\label{fluxes1}
\end{eqnarray}
and it can be shown that particle 4-current, energy momentum tensor etc. are divided into equilibrium and non-equilibrium part as follows,
\begin{eqnarray}
N^\mu(x)&=&N^\mu_{eq}(x)+\delta N^\mu(x)\,,\nonumber\\
\nonumber\\
T^{\mu\nu}(x)&=&T^{\mu\nu}_{eq}(x)+\delta T^{\mu\nu}(x)\,,
\label{fluxes2}
\end{eqnarray}
where $\displaystyle{\delta N^\mu}$ etc. is the non-equilibrium/deviation part from the corresponding quantity.
The entropy 4-current can also be divided into an equilibrium part and an non-equilibrium part as follows,
\begin{equation}
S^\mu(x)=S^\mu_{eq}(x)+\delta S^\mu(x)
\label{entropy1} 
\end{equation}
Now to calculate $\displaystyle{\delta S^\mu}$, we may similarly resort to Grad's 14-moment approximation with $\displaystyle{S^\mu(x)}$ defined as,
\begin{equation}
S^\mu(x)=-\int{dw\,p^\mu\psi(f)},
\label{entropy2}
\end{equation}
where
\begin{equation}
\psi(f)=f(x,p)\ln[A_0^{-1}f(x,p)]-a^{-1}A_0\times\ln\Delta(x,p)
\label{entropy3}
\end{equation}

Now we can expand $\displaystyle{\psi(f)}$ around $\displaystyle{\psi(f^{eq})}$ upto third order in derivative to obtain,
\begin{eqnarray}
\psi(f)&=&-a^{-1}A_0\ln\Delta^{eq}(x,p)+[\alpha(x)-\beta_\nu(x)p^\nu]\nonumber\\
       &\times&[f(x,p)-f^{eq}(x,p)]+\frac{1}{2}[f^{eq}(x,p)A_0^{-1}\nonumber\\
       &\times&\Delta^{eq}(x,p)]^{-1}[f(x,p)-f^{eq}(x,p)]^2\nonumber\\
       &+&\frac{1}{6}[-f^{eq}(x,p)A_0^{-1}\Delta^{eq}(x,p)]^{-2}[f(x,p)-f^{eq}(x,p)]^3\nonumber\\
 \label{entropy4}
\end{eqnarray}

Inserting the above equation in the entropy flux expression i.e. Eq.~\ref{entropy2} we have
\begin{eqnarray}
S^\mu &=& S^\mu_{eq}+\int{dw\,p^\mu[\alpha(x)+\beta_\nu(x)p^\nu](f-f_{eq})}\nonumber\\&-&\frac{1}{2}\int{dw\,p^\mu[f^{eq}(x,p)A_0^{-1}\Delta^{eq}(x,p)]^{-1}}(f-f^{eq})^2\nonumber\\
&-&\frac{1}{6}\int{dw\,p^\mu[f^{eq}(x,p)A_0^{-1}\Delta^{eq}(x,p)]^{-2}}(f-f^{eq})^3\,,\nonumber\\
\label{entropy5}
\end{eqnarray}
where second, third and fourth integration should give us first, second and third terms of entropy current respectively.

A small linear departure function (non-equilibrium) $\displaystyle{\phi(x,p)}$ in $\displaystyle{f(x,p)}$ has been taken as,
\begin{equation}
\phi(x,p)=y(x,p)-y_{eq}(x,p)\approx \epsilon(x)-\epsilon_\mu(x)p^\mu+\epsilon_{\mu\nu}(x)p^\mu p^\nu\,
\label{entropy6}
\end{equation}
where
\begin{equation}
y(x,p)={\rm ln}[A_0^{-1}f(x,p)/\Delta(x,p)]
\label{entropy7}
\end{equation}
differs from its local equilibrium value, $\displaystyle{y_{eq}}$ by quantity $\phi(x,p)$. The moments $\displaystyle{\epsilon,\,\epsilon_\mu\, \text{and}\, \epsilon_{\mu\nu}}$ are assumed small. The expressions for these coefficients are given in the appendix of this paper.

After integration, the entropy 4-current can be written up to third order or cubic in dissipative fluxes as,
\begin{eqnarray}
S^\mu&=&S^0_1u^\mu+S^1_1\Pi u^\mu+S_2^1q^\mu\nonumber\\
&+&\left(S^2_1\Pi^2-S^2_2q^\alpha q_\alpha-S^2_3\pi^{2\langle\alpha\alpha\rangle}\right)\beta u^\mu+\beta\left(S^2_4\Pi q^\mu\right.\nonumber\\
&+&\left. S^2_5\pi^{\langle\mu\alpha\rangle}q_\alpha\right)+\left(S^3_1\Pi^3-S^3_2\Pi q_\alpha q^\alpha+S^3_3\Pi\pi^{2\langle\alpha\alpha\rangle}\right.\nonumber\\
&+&\left. S^3_4q_\alpha q_\beta\pi^{\langle\alpha\beta\rangle}-S^3_5\pi^{3\langle\alpha\alpha\rangle}\right)\beta u^\mu + \left(S^3_6\Pi^2-S^3_7q_\alpha q^\alpha\right.\nonumber\\
&+&\left. S^3_8\pi^{2\langle\alpha\alpha\rangle}\right)\beta q^\mu +\beta\left(S^3_9\Pi\pi^{\langle\mu\alpha\rangle}q_\alpha+S^3_{10}\pi^{2\langle\mu\alpha\rangle}q_\alpha\right)\nonumber\\
\label{entropy8}
\end{eqnarray}
where the coefficients $\displaystyle{S^m_n}$ are calculated as functions functions of $\displaystyle{\varepsilon\,\text{and}\,n}$ and are shown in Appendix A. The superscript in the coefficients denotes the order and the subscript labels the coefficient number in that order. The terms $\displaystyle{\pi^{2\langle\alpha\alpha\rangle}\,=\,\pi^{\langle\alpha\beta\rangle}\pi_{\langle\alpha\beta\rangle}}$, $\displaystyle{\pi^{3\langle\alpha\alpha\rangle}\,=\,\pi^{\langle\alpha\beta\rangle}\pi_{\langle\alpha\delta}\pi^\delta_{\beta\rangle}}$ etc. are written in shortened form. In case of second order theory, $\displaystyle{S^2_n}$ s are equivalent to the coeffcients $\displaystyle{\alpha_i\,\rm{s}\,\rm{and}\,\beta_i\,\rm{s}}$ shown in Ref.~\cite{Muronga:2006zx}. For thermodynamic processes, the entropy principle suggests, $\displaystyle{\partial_\mu S^\mu \geqslant 0}$. The dissipative fluxes can be obtained either from the equations of the balance of the fluxes or from entropy principle. We may recall that at Zeroth order the dissipative fluxes take their equilibrium values, $\displaystyle{\Pi=\Pi_{eq}=0}$, $\displaystyle{q^\alpha=q^\alpha_{eq}=0}$ and $\displaystyle{\pi^{\langle\alpha\beta\rangle}=\pi^{\langle\alpha\beta\rangle}_{eq}=0}$. The complete third order equations of motion or relaxation equations of dissipative fluxes have been given in Appendices A and B.

\subsection*{Bjorken scaling solution}
In the Bjorken scaling solution, the thermodynamic variables such as temperature, chemical potential, pressure and dissipative fluxes are functions of proper time $\displaystyle{\tau} $ only. This means that the explicit transverse space derivatives of temperature, fluid velocity and coupling coefficients etc. are absent. Also under such condition heat flow is shown to be $\displaystyle{q^\mu\,=\,0}$~\cite{Muronga:2003ta,El:2008yy}. However this doesn't imply that the number density $\displaystyle{n}\,=\,0$. In current scenario $\displaystyle{\varepsilon+P}$ is the effective enthalpy, $\displaystyle{P+\Pi+\pi}$ is the longitudinal pressure, and $\displaystyle{P+\Pi-\pi/2}$ is the transverse pressure of the system. 4-velocity in Bjorken (1+1)D expansion is defined as $\displaystyle{u^\mu = (t/\tau,0,0,z/\tau)}$, and the derivatives of 4-velocity are shown to be $\displaystyle{\partial_\mu u^\mu=\frac{1}{\tau}}$ and $\displaystyle{u^\mu\partial_\mu=\frac{\partial}{\partial\tau}}$. In the comoving frame, the shear tensor is diagonal, with a positive shear pressure: $\displaystyle{\pi_{\mu\nu}=\text{diag}(0,\pi/2,\,\pi/2,\,-\pi)}$. The parentheses on the indices denote symmetrization and skew-symmetrization as follows,
\begin{eqnarray}
a^{(\mu\nu)}&\equiv &\frac{1}{2}\left(a^{\mu\nu}+a^{\nu\mu}\right)\nonumber\\
a^{\langle\mu\nu\rangle}&\equiv &(\Delta_\alpha^{(\mu}\Delta_\beta^{\nu)}-\frac{1}{3}\Delta^{\mu\nu}\Delta_{\alpha\beta})a^{\alpha\beta}
\end{eqnarray}

Thus in Bjorken (1+1)D scaling, the first order transport equation could be reduced to 
\begin{eqnarray}
\Pi = -&\zeta &\frac{1}{\tau}\,,\\
\nonumber\\
q^\mu =~&0& \,,\\
\nonumber\\
\pi = \frac{4}{3}&\eta &\frac{1}{\tau}.
\label{viscous1}
\end{eqnarray}

From entropy principle, the general equation in dissipative fluxes can be shown to be,
\begin{eqnarray}
T\partial_\mu S^\mu &=& \frac{\Pi^2}{\zeta}-\frac{q_\mu q^\mu}{\kappa T}+\frac{\pi_{\mu\nu}\pi^{\mu\nu}}{2\eta}\geqslant\,0\,,\nonumber\\
&{\rm with}&\;\; \zeta,\;\kappa,\;\eta\,\geqslant\; 0
\label{entropyrel1}
\end{eqnarray}
where $\displaystyle{\zeta,\;\kappa\;\rm{and}\,\eta}$ are bulk viscous, thermal conductivity, and shear viscous coefficients respectively.

The expressions for shear $\displaystyle{\pi^{\mu\nu}}$, bulk pressures $\displaystyle{\Pi}$ and heat $\displaystyle{q^\mu}$ flux derived from Eqs.~\ref{entropy8} and \ref{entropyrel1} have been shown in Appendix B. In the present calculations where we are working in Bjorken scaling solution, and ultra-relativistic regime (massless fluid particles), only shear pressure equation remains to be solved. The third order shear pressure expression thus obtained as,

\begin{eqnarray}
\pi^{\langle\mu\nu\rangle}&=&-2\eta T\Delta^{\alpha\mu}\Delta^{\beta\nu}[-\partial_{\langle\alpha}u_{\beta\rangle}+2\beta S^2_3(u_\lambda\partial^\lambda\pi_{\langle\alpha\beta\rangle})\nonumber\\
&+&\pi_{\langle\alpha\beta\rangle}\partial^\lambda(\beta u_\lambda S^2_3)
+ 3\beta S^3_5 (u_\lambda\partial^\lambda\pi_{\langle\alpha\delta})\pi^{\delta}_{\beta\rangle}\nonumber\\
&+&\pi_{\langle\alpha\delta}\pi^{\delta}_{\beta\rangle}\partial^\lambda(\beta S^3_5u_\lambda)]\,,
\label{viscous3}
\end{eqnarray}

Using Eq.~\ref{viscous3}, third order equation can also be written as,

\begin{eqnarray}
u_\lambda\partial^\lambda\pi^{\langle\mu\nu\rangle}= &-&\frac{\pi^{\langle\mu\nu\rangle}}{\tau_\pi}+\frac{2\eta\,\partial^\mu u^\nu}{\tau_\pi}-\frac{2\eta T}{\tau_\pi}\pi^{\langle\mu\nu\rangle}\partial^\lambda\left(\frac{S^2_3}{2T}u_\lambda\right)\nonumber\\
&-& \frac{3\eta S^3_5}{\tau_\pi}(u_\lambda\partial^\lambda\pi^{\langle\mu\delta})\pi_{\delta}^{\nu\rangle}\nonumber\\ &-& \frac{2\eta T}{\tau_\pi}\pi^{\langle\mu\delta}\pi_{\delta}^{\nu\rangle}\partial^\lambda\left(\frac{S^3_5}{2T}u_\lambda\right)\,.
\label{viscous4}
\end{eqnarray}
where $\displaystyle{\tau_\pi=2\eta S^2_3}$ is relaxation time for the shear pressure. The coefficient, $\displaystyle{S^2_3}$ is taken to be $\sim 9/4\varepsilon$ in the ultra-relativistic limits. We have used equation of state (EoS) due to assumed ultra-relativistic scenario to be, $\displaystyle{\varepsilon=3P}$. 

After simplification and keeping all the terms, the final equations for shear pressure for 3$^{\rm rd}$ order viscous and massless fluids is found to be 

\begin{eqnarray}
\dot{\pi}&=&-\frac{\pi}{\tau_\pi}-\frac{1}{2}\frac{\pi}{\tau}+\frac{3}{10}\frac{\varepsilon}{\tau}+\frac{5}{8}\frac{\pi}{\varepsilon}\dot{\varepsilon}-\frac{3}{2}\frac{\pi^2}{\varepsilon\tau}+\frac{27}{8}\frac{\pi^2}{\varepsilon^2}\dot{\varepsilon}-\frac{12}{5}\frac{\pi}{\varepsilon}\dot{\pi}\nonumber\\
\nonumber\\
\label{viscous5}
\end{eqnarray}

Also in (1+1)D Bjorken flow, the energy and number density equations calculated from Eqs.~\ref{number} and~\ref{energymomentumtensor} are similarly given by,
\begin{equation}
\dot{\varepsilon}=-\frac{\varepsilon+P}{\tau}+\frac{\pi}{\tau},\;\;\dot{n}=-\frac{n}{\tau}\,.
\label{energyequation}
\end{equation}

The shear differential equation shown in Eq.~\ref{viscous4} has order by order implication on the final output or calculated energy and entropy densities. The effects due to inclusion of various orders on the solutions of dissipative equations will be discussed next section.
%
%

\section{Results and Discussions.}


\begin{figure}[ht]
\includegraphics[scale=0.3]{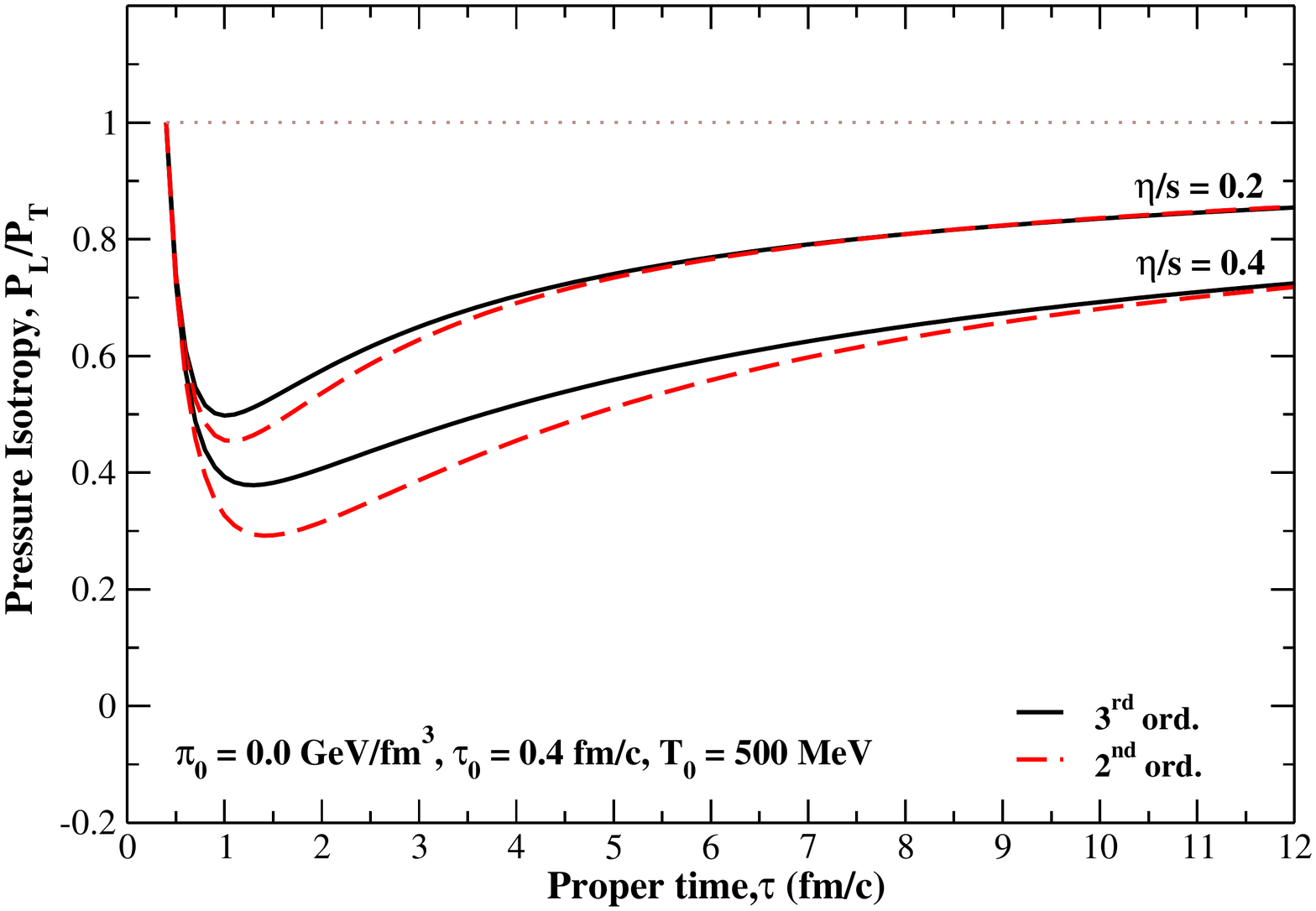}
\caption{\label{figPI}(Colour online) Pressure isotropy of relativistic fluid using second and third order shear equations with $\displaystyle{\pi_0}$ = 0.}
\end{figure}

Pressure isotropy is the measure of system isotropization and indicates system returning to equilibrium. It is calculated as the ratio of longitudinal pressure ($P_L$) to transverse pressure ($P_T$) of the system. A system of quarks and gluons in equilibrium behaves like a fireball where $P_L$ equals $P_T$, and the ratio remains unity and independent of time. But with the presence of viscous forces, the ratio diverges from unity (i.e. the system is out-of-equilibrium) but gradually returns to equilibrium with time. Usually the collision axis for the heavy ions are taken to be longitudinal axis ($viz.$ z-axis) with zero momentum along 'x' or 'y' (transverse) directions. After the collision because of the presence of viscous drag along the z-axis, the particles tend to push more on the transverse direction than the longitudinal direction. However the system due to particle interaction and decreasing dissipative fluxes, the pressure continuous to build along z-axis and if given enough time the system could return to equal pressures along both longitudinal and transverse axes. Let us now move to discussion of the results. We may recall that calculations have been done in (1+1)D Bjorken scenario with boost-invariance assumed along z-axis.

Fig.~\ref{figPI} shows pressure isotropy  as function of proper-time for an expanding fluid with initial condition matching that of ideal relativistic fluid. The figure shows the comparison between shear pressure in 2$^{\rm nd}$ and 3$^{\rm rd}$ order equations. Two different values of $\displaystyle{\eta/s}$ have been used. Because of the initial $\pi_0=0$ GeV/fm$^\text{3}$ the fluid is initially isotropic and thereafter the dissipative fluxes put the system rapidly out of equilibrium. The system tries to get back to equilibration and the trends show a continuous rise in isotropization although a certain degree of saturation already sets after 4-5 fm/c. The ideal scenario is represented by unit valued line (dotted). Initially second order shows a greater dip than the third order, which indicates that third order tends to limit or decrease the dissipative effects brought in by lower orders. One may also find that higher value of $\displaystyle{\eta/s}$ bring in more difference between 2$^{\rm nd}$ and 3$^{\rm rd}$ order shear equations. An initial temperature  of $T_0$ = 500 MeV at QGP formation time of $\tau_0$=0.4 fm/c has been used in the calculations. It is interesting to mention that in some earlier studies higher orders beyond third have been considered by A. El et al. and the seminal paper has heuristically explained that a higher order beyond third might increase the shear effects and observables might be closer to second order. However it is also suggested in that literature that the effects might be oscillatory in nature if more ordered corrections are taken in the picture. This is yet to be studied more extensively~\cite{El:2009vj}.

\begin{figure}[h]
\includegraphics[scale=0.3]{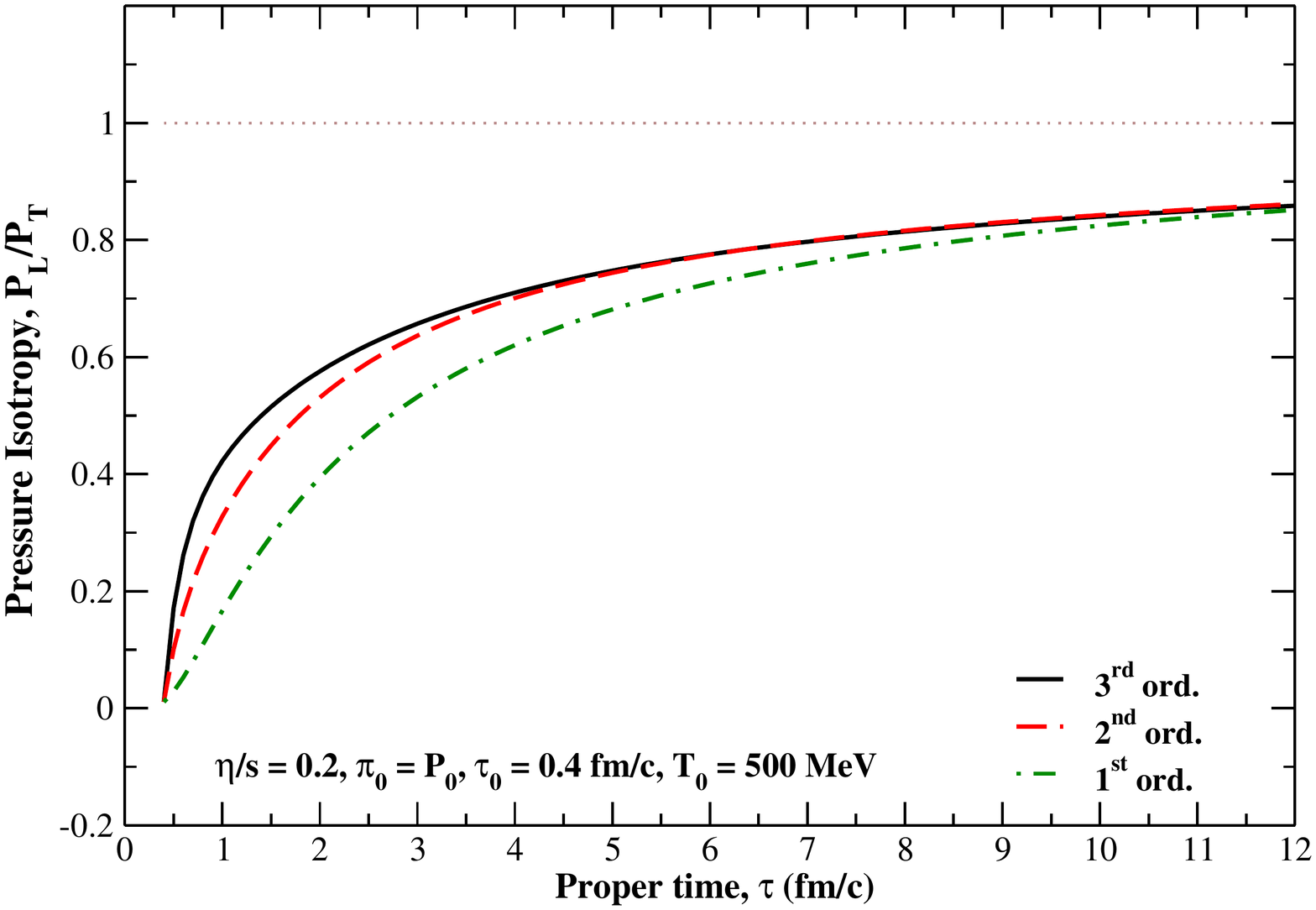}
\caption{\label{figPINS}(Colour online) Pressure isotropy of relativistic fluid using second and third order shear equations with $\displaystyle{\pi_0}\,=\,P_0$.}
\end{figure}

Fig.~\ref{figPINS} shows pressure isotropy when initial shear pressure is changed from ideal fluid to $\displaystyle{\pi_0\,=\,P_0}$. We also recall that while deriving transport equations, we assumed that the dissipative fluxes should be small as compared to the primary physical quantitites ($\displaystyle{\varepsilon,\,n,\,P}$). In terms of $\displaystyle{\pi}$ this condition can be written as $\displaystyle{(\pi^{\mu\nu}\pi_{\mu\nu})^{1/2}\,<\,P}$. In general the thermodynamic quantities will decrease with time as long as the condition $\displaystyle{\pi\,\ll\,\varepsilon+P}$, which in our case is $\displaystyle{\pi_0\,\ll\,4P_0}$. A moderate value of $\displaystyle{\pi_0\,\leq\,P_0}$ has been used to avoid negative effective enthalpy and negative longitudinal pressure. Unlike the first order theories such limitations could be put on initial conditions in both second and third order theories. But for investigative studies one may relax this binding on $\displaystyle{\pi_0}$ to $\displaystyle{4P_0}$. The quantity $\displaystyle{P_L/P_T}$ rises rapidly for the second and third order equations similar to previous figure but the rates decrease and the curves try to merge beyond 6 fm/c. The first order result starts slower but tends to merge around 12 fm/c. Here too, inclusion of third order decreases the effects of dissipative fluxes when compared to second order and first order shears. The initial temperature and time identical to Fig.~\ref{figPI}, and a modest value of $\displaystyle{\eta/s}\,=\,0.2$ have been used for this plot.

Next we move over to comparison between various models which incorporated third order viscous hydrodynamics.

\begin{figure}[h]
\includegraphics[scale=0.3]{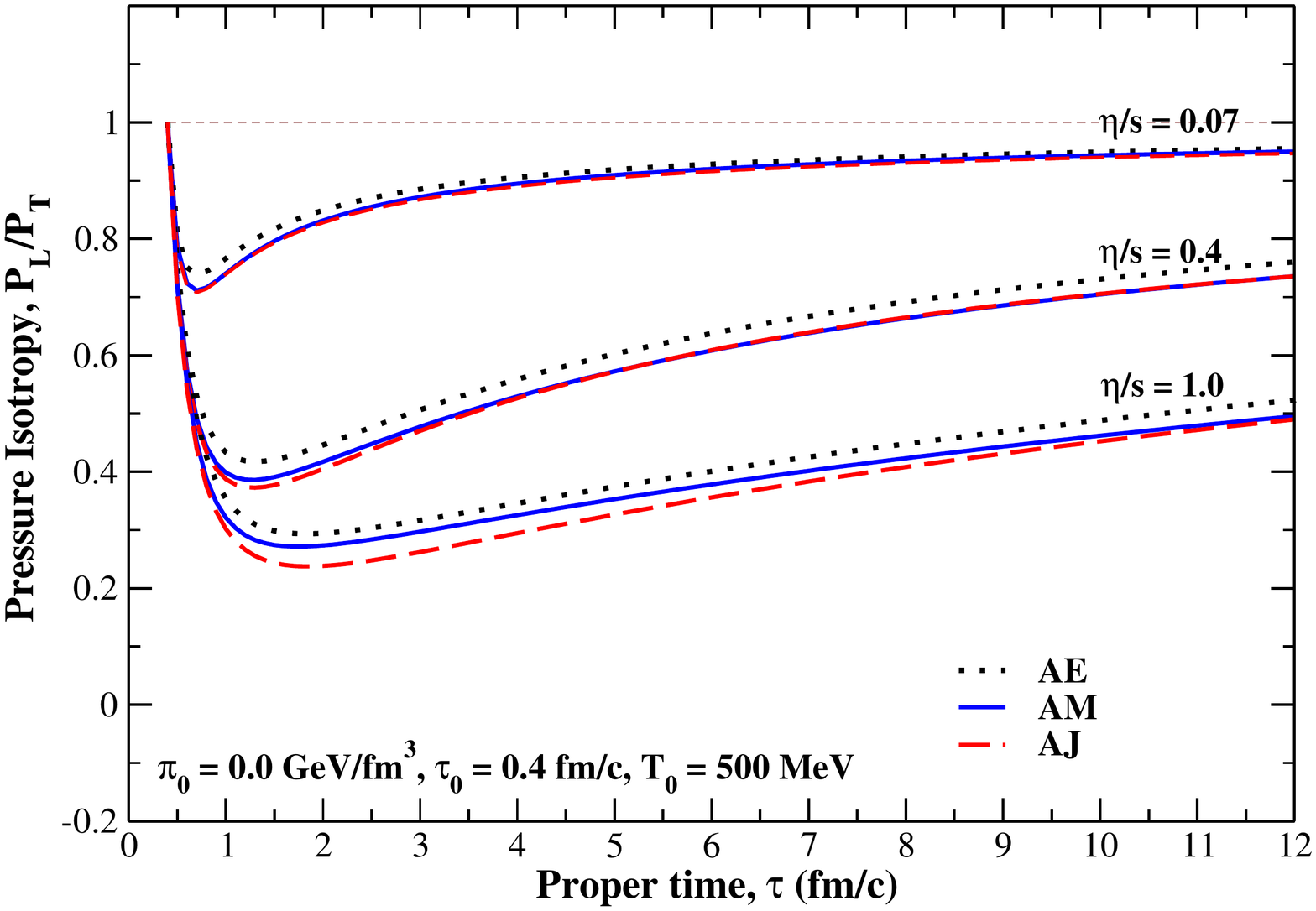}
\caption{\label{figPIMC}(Colour online) Comparison of models on time evolution of pressure isotropization for different $\displaystyle{\eta/s}$ values. A. Muronga et al. (AM) denotes inclusion of $\displaystyle{\partial_iS^2_3~\text{and}~\partial_iS^3_5}$ terms in the third order equation.}
\end{figure} 

\begin{figure}[h]
\includegraphics[scale=0.3]{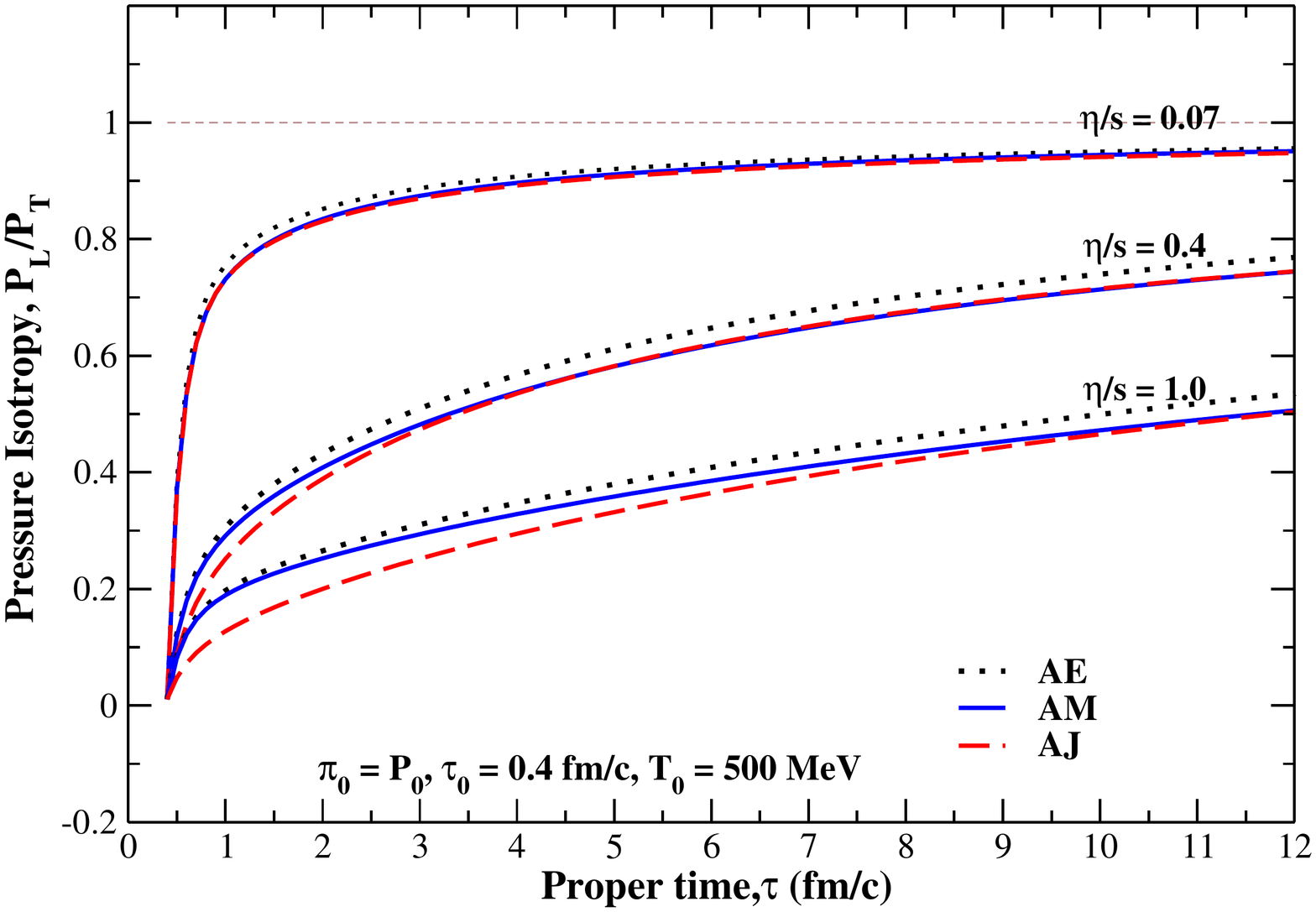}
\caption{\label{figPIMCNS}(Colour online) Comparison of models on time evolution of pressure isotropization for different $\displaystyle{\eta/s}$ for $\displaystyle{\pi_0}$=$\displaystyle{4\eta/3\tau_0}$.}
\end{figure}

\begin{figure*}[ht]
\includegraphics[scale=0.3]{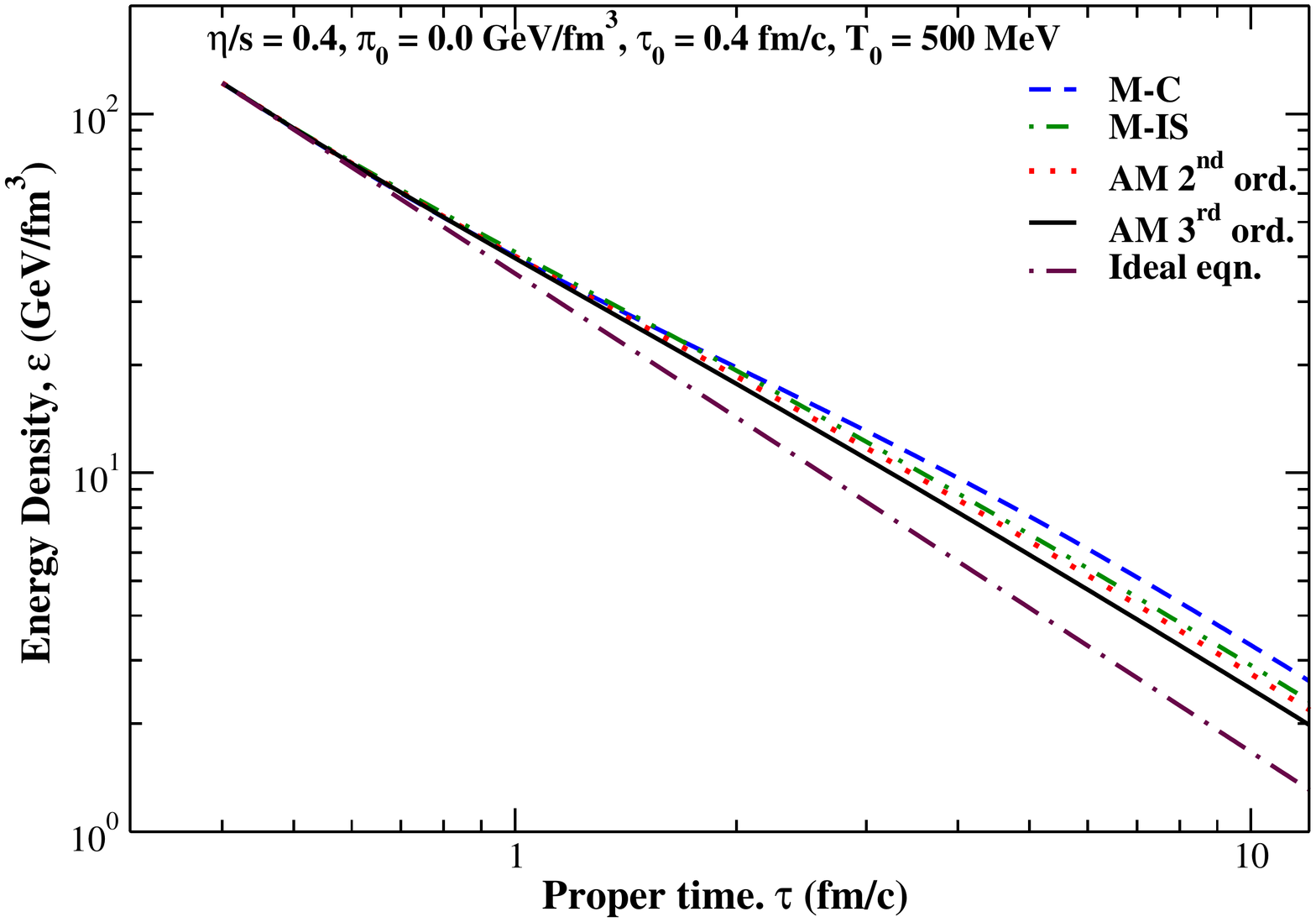}
\includegraphics[scale=0.3]{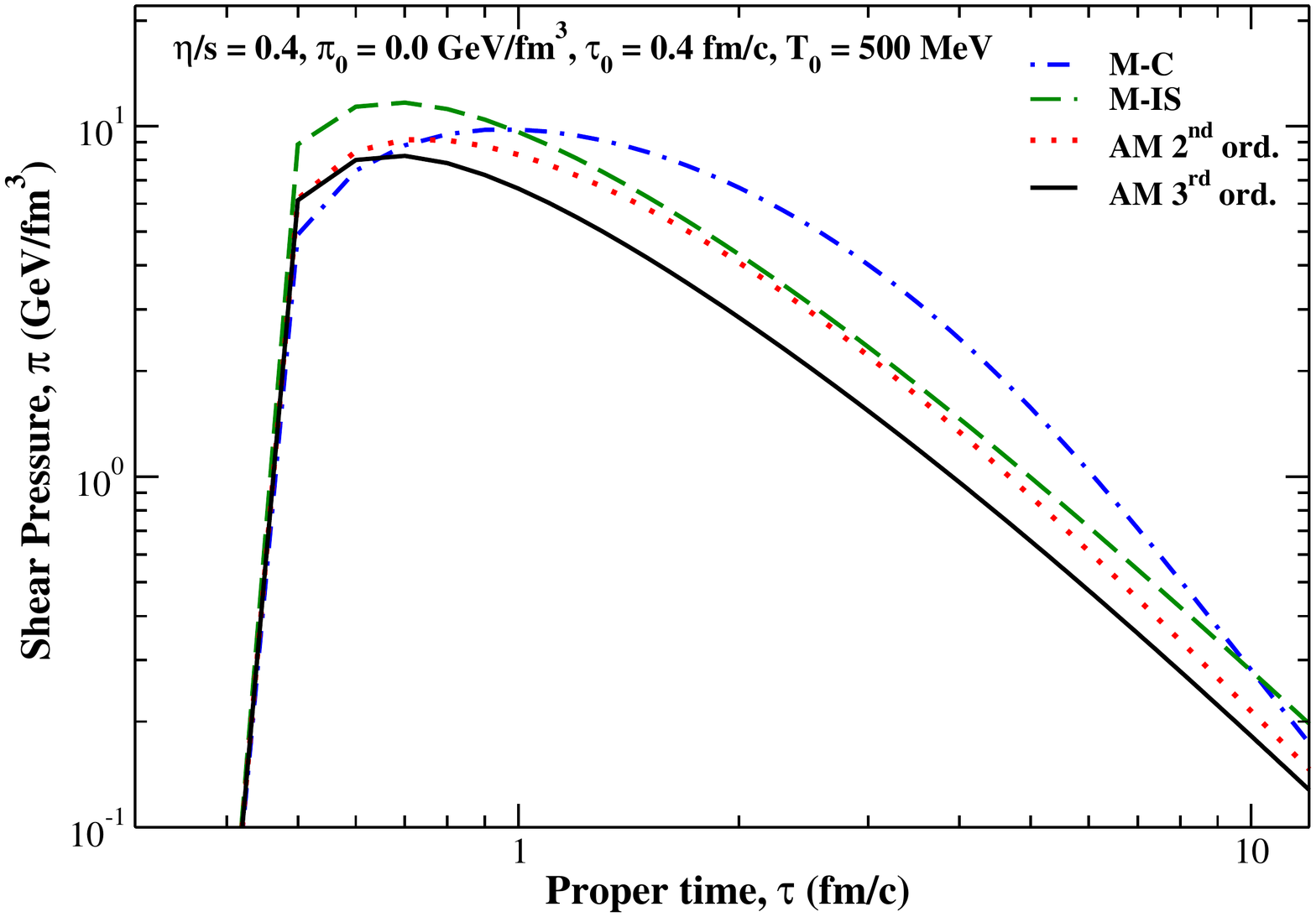}
\caption{\label{figwide}(Colour Online) (Left) Energy density solution for maxwell-Cattaneo like equations, Muller-Israel-Stewart theory and A. Muronga et al. 2$^{nd}$ and 3$^{rd}$ order equations, (Right) Solutions for shear pressure for models mentioned.}
\end{figure*}

\begin{figure}[h]
\includegraphics[scale=0.3]{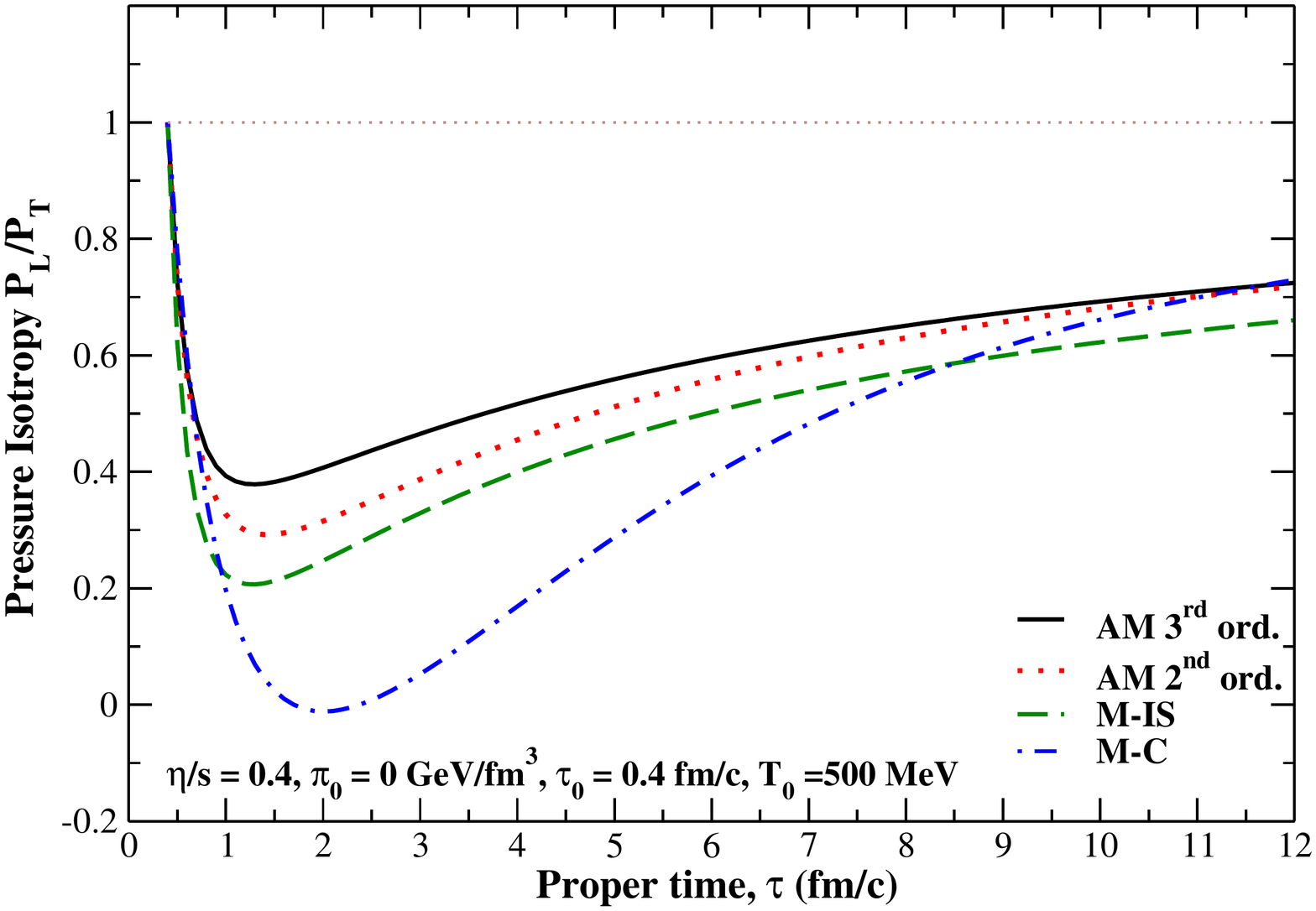}
\caption{\label{figPIterm}(Colour online) Time evolution of pressure isotropy for various terms in shear pressure differential equation.}
\end{figure}

To begin with Figs.~\ref{figPIMC} and~\ref{figPIMCNS}, it can be stated that values of the coefficients in relaxation equations from models differ from each other respectively. The models used for comaprison are referred here as A. El et al. (AE), a third order thermodynamic theory, A. Jaiswal et al. (AJ), a third order kinetic theory, and current work, A. Muronga et al. (AM) which is based on third order theory based on thermodynamic entropy principle. The differential equations in these models have been found to be very sensitive to the values of coupling coefficients. One may also find that certain terms have been neglected in Muller-Israel-Stewart theories for second order because of their non linearity. A. Muronga et al. model has included these terms in developing the shear differential equations.
The figures Figs.~\ref{figPIMC} and~\ref{figPIMCNS} aim to highlight the sensitivity of values of the coefficients and terms used in the models. The model by AM calculates full expansion of entropy four current using Grad's 14-moments theory. As approximations,  mass of fluid particles and heat flux have been neglected in order to obtain a very simple picture of how viscous drag forces work within the system. Also compared to AE model, terms of order 4 in $\displaystyle{\pi/\varepsilon}$ have also been included in AM model. As seen from the figure that the models do not differ much for for lower $\displaystyle{\eta/s}$ values. For higher values of the parameter, the models differ by a small magnitude at low $\displaystyle{\tau}$ while for at latter times, they tend to merge. The Figs.~\ref{figPIMC} and~\ref{figPIMCNS} also differ from each other at the starting point of the curve due to choice of initial values of $\displaystyle{\pi_0}$. However the evolution of pressure isotropy ratios show similar trend after $\displaystyle{\tau}$ = 2 fm/c in both figures.
The effects of various terms in Eq.~\ref{viscous4} (simplified into Eq.~\ref{viscous5}) will be discussed for Figs.~\ref{figwide} and~\ref{figPIterm}.

\begin{figure}[h]
\includegraphics[scale=0.3]{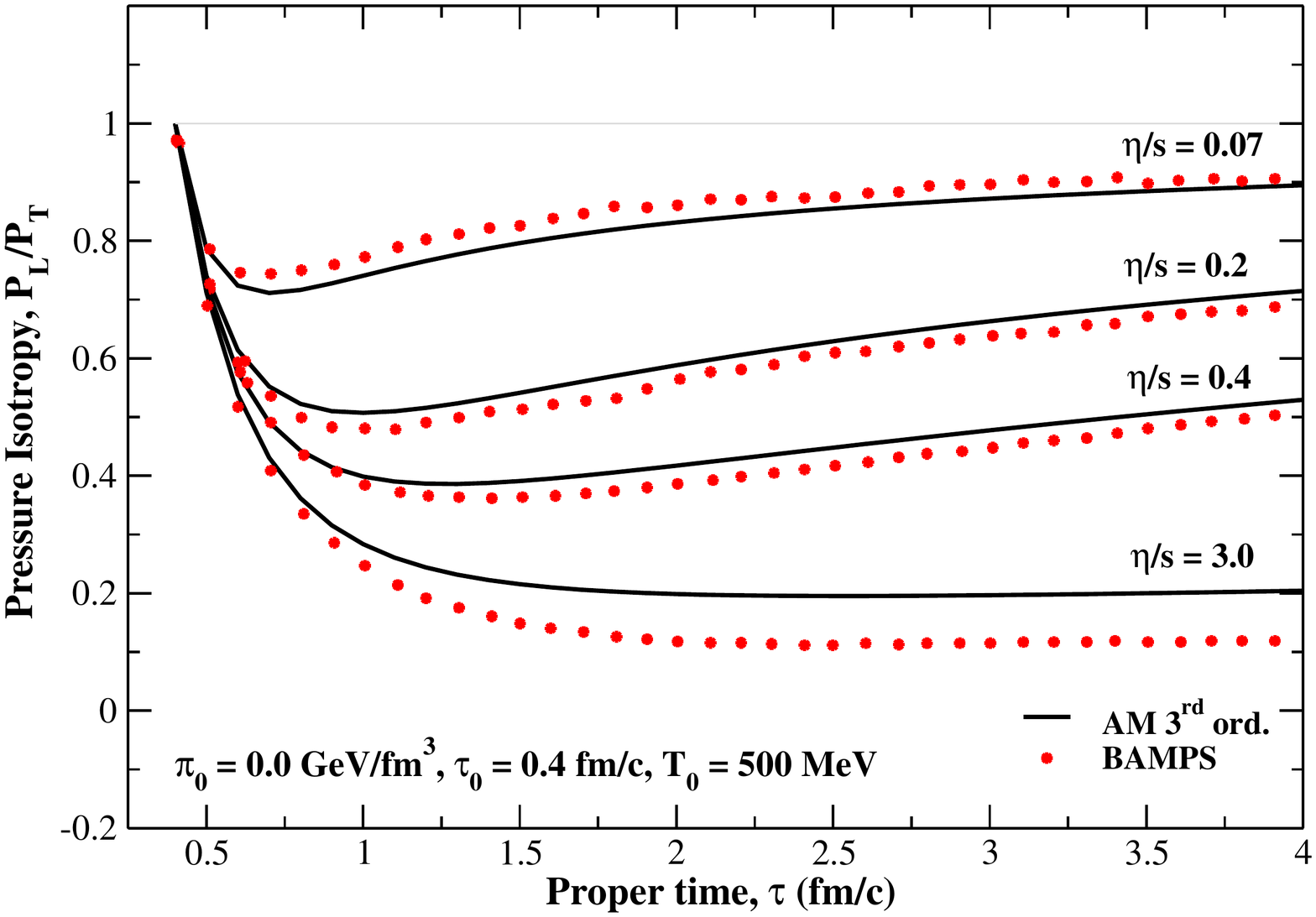}
\caption{\label{figBAMPS}(Colour online) Comparison pressure isotropy ration from AM third order theory with BAMPS transport calculation for different $\displaystyle{\eta/s}$.}
\end{figure}

In Fig.~\ref{figBAMPS} the pressure issotropy ratio $\displaystyle{P_L/P_T}$ calculated from current model (AM) has been compared to results from BAMPS transport theory. BAMPS data points have been extracted from A. El et al.'s paper which has upto $\tau$ = 4 fm/c and hence the figure has shorter x-axis range compared to other figures. The main focus of this plot is to highlight the difference between transport calculation and effective third order theory shown in the current work. The BAMPS exhibit larger transverse pressure compared to AM model for $\displaystyle{\eta/s}$ values $\displaystyle{\geqslant}$ 0.2. The ratio however shows similar trend for both models as system evolves with time.

In Figs.~\ref{figwide} and~\ref{figPIterm}, Eq.~\ref{viscous4} has been dissected and discussed term by term. If Eq.~\ref{viscous4} is simplified and the first two terms on r.h.s. of the equation are considerd, they give us Maxwell-Cattaneo (M-C) like equation. As known from earlier studies that M-C equation has been used to study the propagation of second sound in presence of dissipative heat flux in materials~\cite{maxwell}. One may draw analogy and write,
\begin{equation}
\frac{dq}{dt}=-\frac{q}{t_0}+k\frac{\Theta}{t_0}\;\Longleftrightarrow\;\frac{d\pi}{d\tau}=-\frac{\pi}{\tau_\pi}+\frac{2\eta}{\tau}\frac{1}{\tau_\pi}.
\label{maxcat}
\end{equation}
Where $\displaystyle{t_0}$ is the relaxation time, $\displaystyle{\Theta}$ is absolute temperature etc. We have labeled the solutions to above set of terms in AM equation as M-C in the figures. In addition we have also compared the solutions of Muller-Israel Stewart equations (M-IS) second order theory. One may recall that certain terms such as $\displaystyle{\Pi\partial_i\varepsilon,\,\pi_{\langle\alpha\beta\rangle}\partial_i\varepsilon,\,q_\alpha\partial_in}$ etc. and consequently $\displaystyle{\partial_iS^2_3\,\text{and}\,\partial_iS^3_5}$ etc. have been neglected in M-IS equations due to their non-linear nature~\cite{Israel:1979wp}.  MI-S theory developed for second order shows that these terms (including vorticity terms) may not be explicit from entropy principle or kinetic theory~\cite{Betz:2009zz}. However the extra terms are consistent with conformal theories~\cite{Baier:2007ix,Koide:2006ef}.

It can be seen from the figures that the terms $\displaystyle{\partial_iS^2_3\,\text{and}\,\partial_iS^3_5}$ have discernible effects on solutions. The calculations are done at a moderate $\displaystyle{\eta/s}$ = 0.4. The differences in the solutions are more visible at low $\displaystyle{\tau}$. The M-C equation gives more shear pressure and energy density (see Fig~\ref{figwide}) for low and intermediate time, and consequently the pressure isotropy parameter, $\displaystyle{P_L/P_T}$ calculated from M-C equation, (see Fig.~\ref{figPIterm}) goes below zero for $\displaystyle{\tau}$ $<$ 3 fm/c. MI-S equations which have 2$\displaystyle{\rm{nd}}$ order shear terms doesn't contain $\displaystyle{\partial_iS^2_3}$ term. As result the solutions to M-IS equation although closer to AM second order theory, have considerable differences with it and thus the effects of non-linear terms could be highlighted. AM second order theory decreases the overall dissipative effects of the shear. AM third order theory shows further decrease in shear pressure and almost merges with second order solution around $\displaystyle{\tau}$ = 12 fm/c. Third order energy density solution also differs from ideal solution by a magnitude. This difference however may be due to assumed value of $\displaystyle{\eta/s}$ = 0.4. The effects of the $\displaystyle{\eta/s}$ on ordered theories have already been discussed in context of Fig.~\ref{figPI}.

\begin{figure}[h]
\includegraphics[scale=0.3]{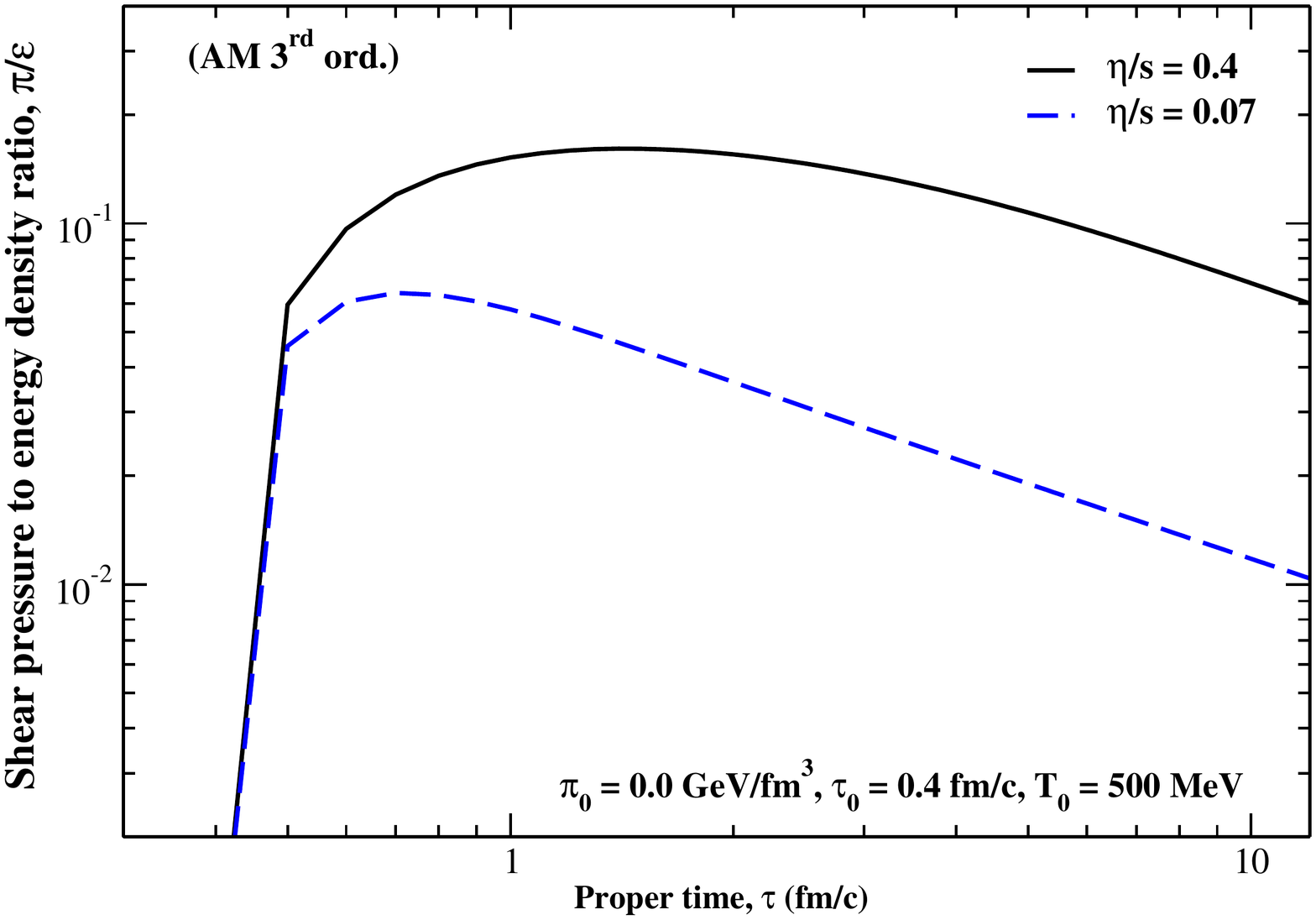}
\caption{\label{figpibye}(Colour online) Time evolution of pressure isotropy for various $\displaystyle{\eta/s}$ values at $T_0$ = 500 MeV and $\tau_0$ = 0.4 fm/c.}
\end{figure}

In Eq.~\ref{viscous4} the term $\displaystyle{-\frac{3\eta S^3_5}{\tau_\pi}u_\lambda(\partial^\lambda\pi^{\mu\delta})\pi_{\delta}^{\nu}}$ appearing in third order shear relaxation equation is non-linear and has been simplified into the last term in Eq.~\ref{viscous5}. The ratio $\displaystyle{\frac{3\eta S^3_5}{\tau_\pi}\pi\,\approx\,\frac{\pi}{\varepsilon}}$ has been neglected in Ref.~\cite{El:2009vj} for all time. However for earlier times when $\displaystyle{\tau\,<\,\tau_\pi}$, this term proves to be relevant. This is evident in Fig.~\ref{figpibye} where for a moderate value of $\displaystyle{\frac{\eta}{s}}$ = 0.4, calculation of the ratio $\displaystyle{\frac{\pi}{\varepsilon}}$ has approximate values between 0.161 and 0.11 at $\displaystyle{\tau}$ = 1.0 fm/c to 7 fm/c but drops down to 0.060 around $\displaystyle{\tau}$ = 12 fm/c. Hence for major part of QGP evolution this term although non-linear becomes a contributing factor. 

\begin{figure}[h]
\includegraphics[scale=0.3]{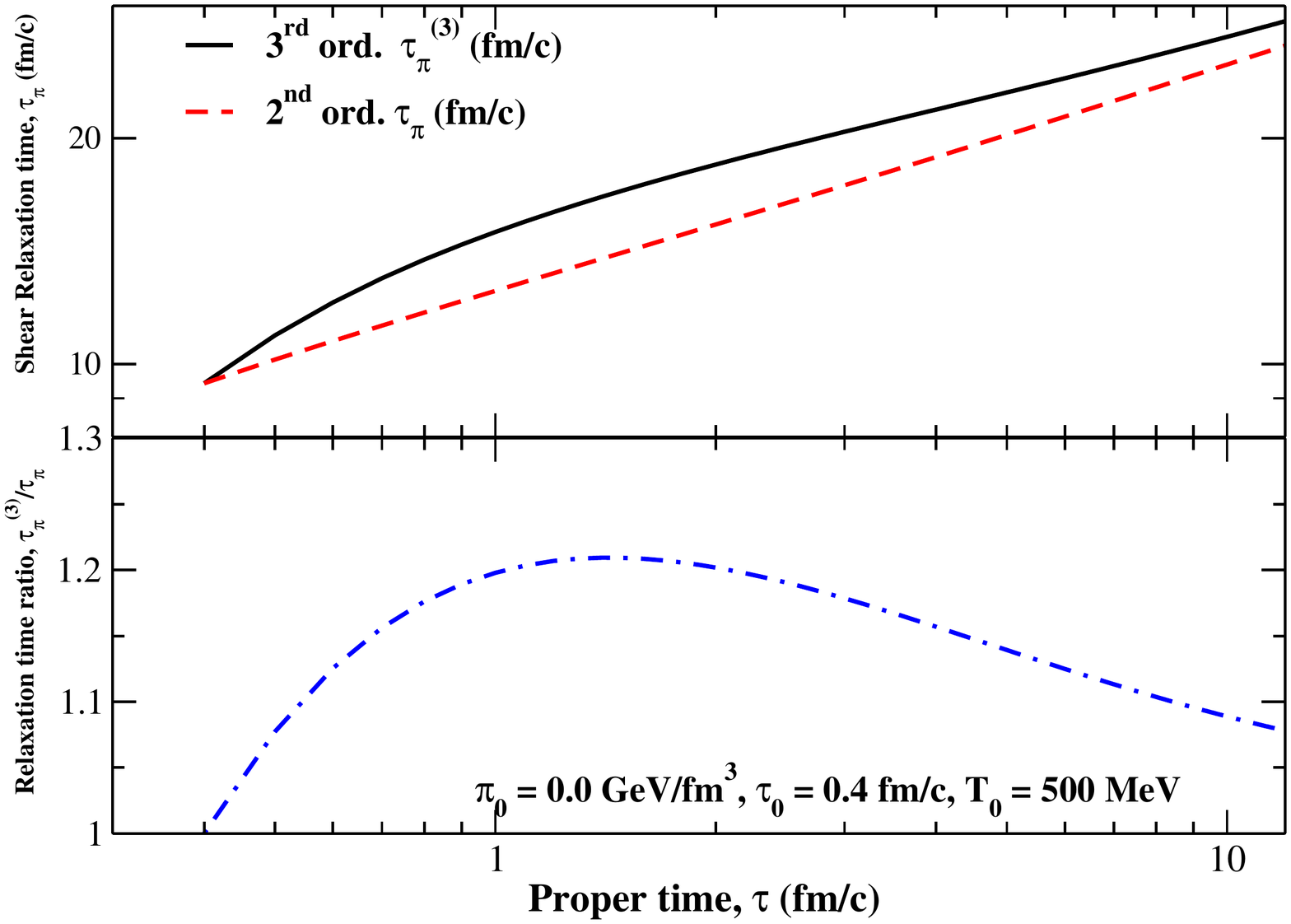}
\caption{\label{figrelaxtime}(Colour online) Time evolution of shear relaxation time for $\displaystyle{\eta/s\,=\,0.4}$ values at $T_0$ = 500 MeV and $\tau_0$ = 0.4 fm/c.}
\end{figure}

Following the above discussion, the term $\displaystyle{3\eta S^3_5\,.\pi}$ can be referred as a relaxation time that comes from third order equation. This term also provides a correction factor to second order relaxation time $\displaystyle{\tau_\pi}$ in Eq.~\ref{viscous4} as follows
\begin{eqnarray}
&\tau_\pi&\left(g^{\mu\delta} +\frac{3\eta S^3_5}{\tau_\pi}\pi^{\langle\mu\delta}\right)u_\lambda\partial^\lambda \pi_{\delta}^{\nu\rangle}\nonumber\\ 
&=& -\pi^{\langle\mu\nu\rangle}+2\eta\,\partial^\mu u^\nu 
- 2\eta T\pi^{\langle\mu\nu\rangle}\partial^\lambda\left(\frac{S^2_3}{2T}u_\lambda\right)\nonumber\\
&-& 2\eta T\pi^{\langle\mu\delta}\pi_{\delta}^{\nu\rangle}\partial^\lambda\left(\frac{S^3_5}{2T}u_\lambda\right)\,,\nonumber\\
\rm{or}\,\,&\tau_\pi^{\langle\mu\delta}&u_\lambda\partial^\lambda\pi_{\delta}^{\nu\rangle}\nonumber\\
&=& -\pi^{\langle\mu\nu\rangle}+2\eta\,\partial^\mu u^\nu - 2\eta T\pi^{\langle\mu\nu\rangle}\partial^\lambda\left(\frac{S^2_3}{2T}u_\lambda\right)\nonumber\\
&-& 2\eta T\pi^{\langle\mu\delta}\pi_{\delta}^{\nu\rangle}\partial^\lambda\left(\frac{S^3_5}{2T}u_\lambda\right).\nonumber\\
\end{eqnarray}
 Where $\displaystyle{\tau_\pi^{\mu\delta}}\,=\,\tau_\pi\left(g^{\mu\delta} +\frac{3\eta S^3_5}{\tau_\pi}\pi^{\mu\delta}\right)$ can be termed as the modified relaxation time for the third order shear viscous pressure. $\displaystyle{\tau_\pi^{\mu\delta}}$ is in a tensorial form and depends on the form of the shear pressure tensor. We find that third order correction explicitly brings in the shear fluxes in the expression for the relaxation time. This was absent in second order. Another interesting feature is that no direction dependent effects were present in second order relaxation time whereas in third order theory we have $\displaystyle{\tau_\pi^{xx},\,\,\tau_\pi^{yy}\,\,\text{and}\,\,\tau_\pi^{zz}}$ along x, y, and z-directions. In the case of Bjorken (1+1)D scenario, the effect of the correction term for relaxation time has been obtained from Eq.~\ref{viscous5} and the modified relaxation time, $\displaystyle{\tau_\pi^{zz}\,=\,\tau_\pi^{(3)}}$ is shown in Fig.~\ref{figrelaxtime}. The figure shows a clear difference between second and third order effects. Thus it would be interesting to study the effects of various ordered theories on the relaxation times.

\begin{figure}[h]
\includegraphics[scale=0.3]{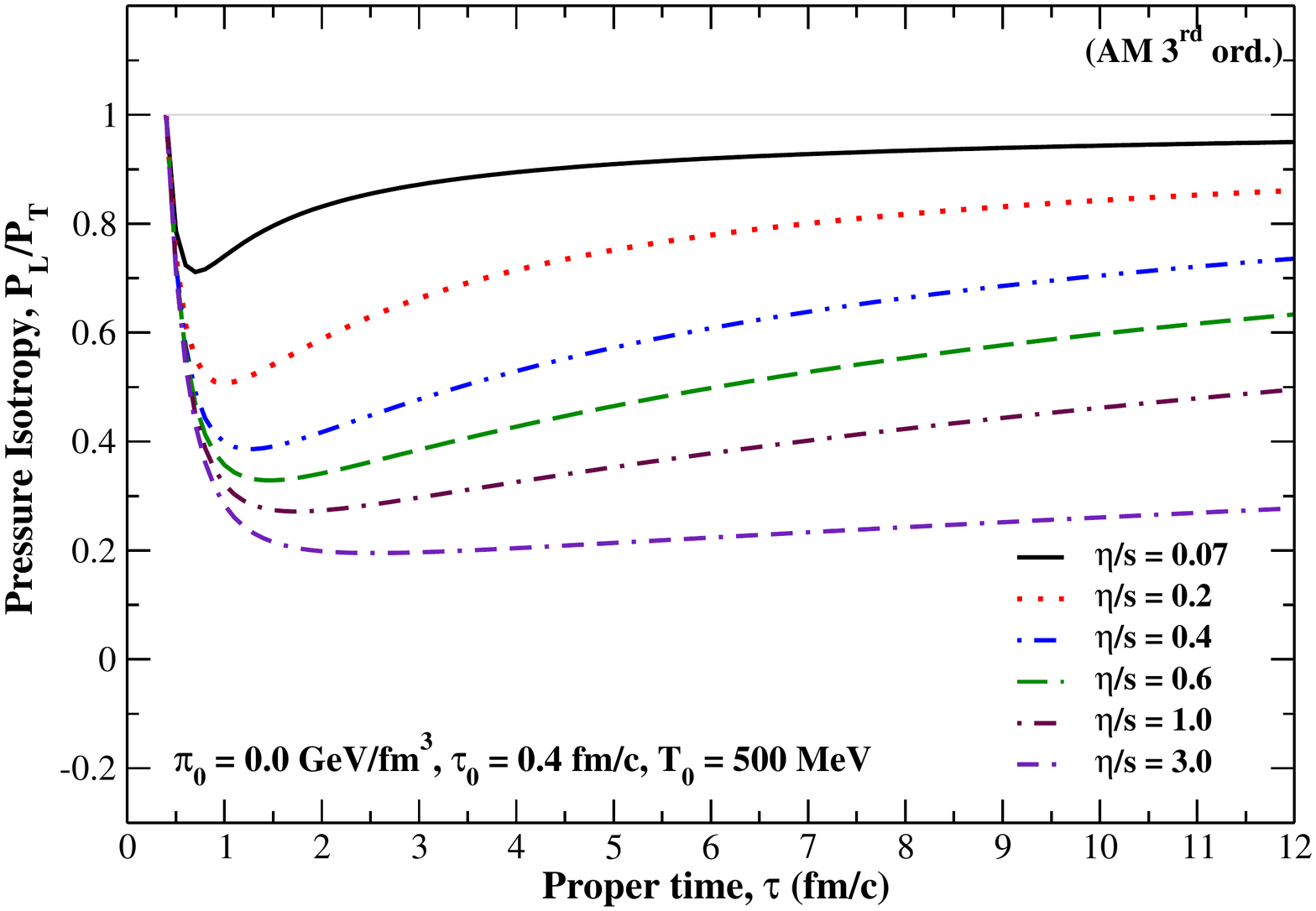}
\caption{\label{figPIEtaS}(Colour online) Time evolution of pressure isotropy for various $\displaystyle{\eta/s}$ values at $T_0$ = 500 MeV and $\tau_0$ = 0.4 fm/c.}
\end{figure}

Let us now discuss the effects of shear viscosity to entropy ratio on the pressure isotropization. It can be shown from transport models that parameter $\displaystyle{\eta/s}$ shows coupling between medium particles or strength of particles' interaction. The parameter could be shown to be $\displaystyle{\eta/s\;\sim\;0.066*\left[\alpha_s^2{\rm ln}(\alpha_s^{-1})\right]^{-1}}$ ($N_f\,=\,3\;{\rm and}\,0\,\ll\,\alpha_s\,\ll\,1$)~\cite{Baym:1990uj,Baym:1997gq}. The equation shows viscosity to entropy ratio as a function of strong coupling. In current calculation, $\displaystyle{\eta/s}$ has been treated as a parameter with constant values which also indicate constant values for strong coupling assumed in our calculations. However it should be recalled that strong coupling is a running coupling and depends on the system temperature or momentum transfer during particle interaction.

Fig.~\ref{figPIEtaS} shows dependence of shear viscosity to entropy ratio on pressure isotropization. As $\displaystyle{\eta/s}$ is increased, the system is removed further from equilibrium and at high values of $\displaystyle{\eta/s}$ = 1.0 - 3.0 (highly viscous fluid), the ratio $\displaystyle{P_L/P_T}$ is almost flat after 2 fm/c with a slow rise. This indicates that fluid with high viscosity may not return to isotropy quickly. The earlier studies~\cite{El:2008yy} have however shown that second order theories breakdown beyond~$\displaystyle{\eta/s\sim 0.4\,-\,0.5}$, possibly limiting the maximum values for $\displaystyle{\eta/s}$.

For a very low value of $\displaystyle{\eta/s}$ = 0.07, the pressure isotropy ratio goes closer to unity around the time $\tau$ = 10 - 12 fm/c. This suggests the importance for the study of time or length scale dependence of isotropization or equilibration of viscous fluids.

\begin{figure}[h]
\includegraphics[scale=0.3]{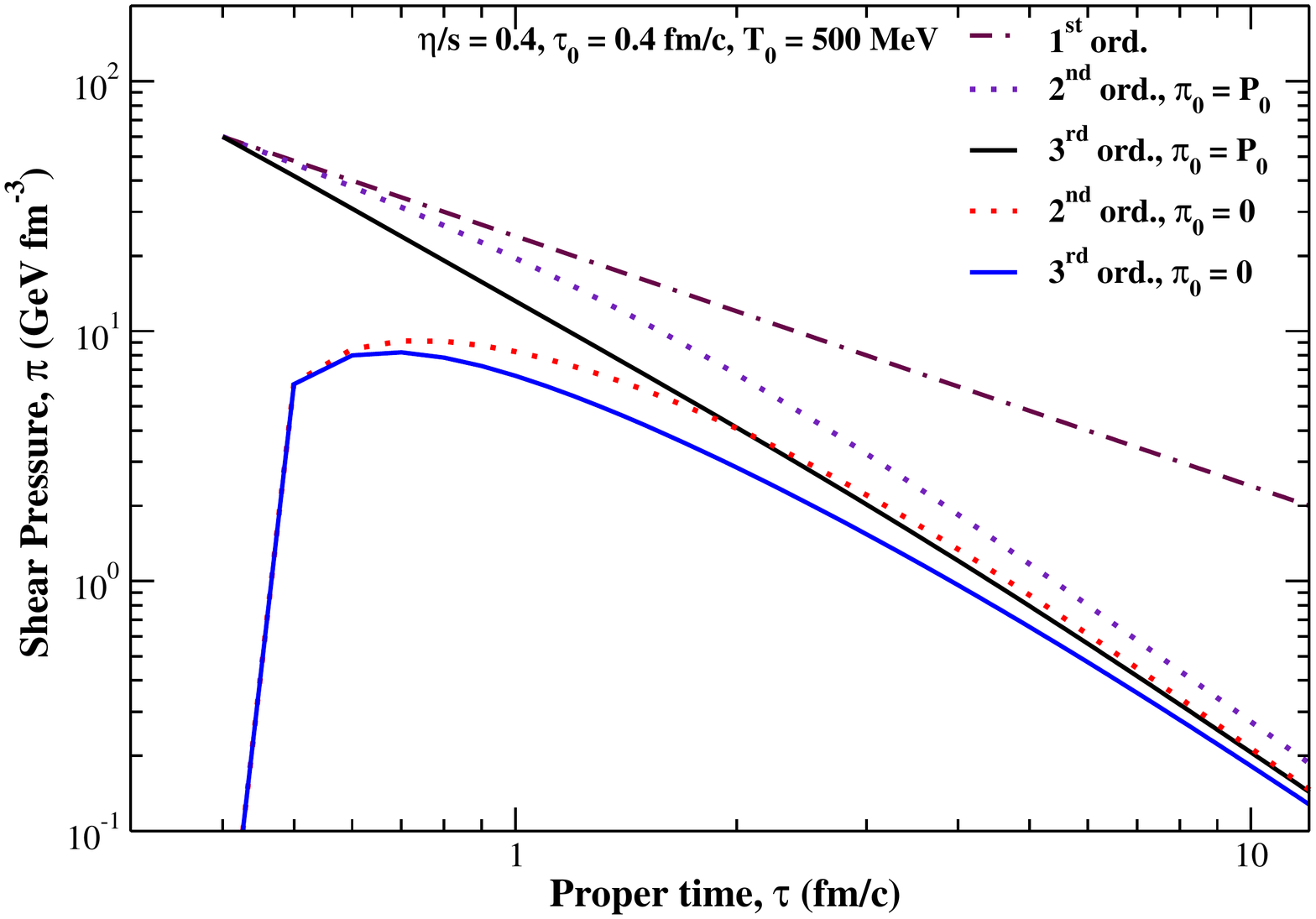}
\caption{\label{figSP}(Colour online) Comparison of different orders of shear pressure for two initial conditions of $\displaystyle{\pi_0}$=0.0 and $\displaystyle{\pi_0}$=$\displaystyle{ {\rm P_0}}$}
\end{figure}

Fig.~\ref{figSP} shows proper-time evolution of shear pressure with two different initial shear pressure. $\displaystyle{\pi_0}$=0 matches that of ideal fluid, while $\displaystyle{\pi_0}$=$\displaystyle{{\rm P_0}}$ provides a high initial dissipative flux. The results from second and third order equations for an ideal fluid like initial condition, rise rapidly from zero but tend to fall off more quickly when compared to the corresponding results using a large initial shear. Irrespective of the initial conditions both second and third order theories bring down the dissipative fluxes as compared to the first order theory. Also third order theory brings shear viscosity to lower value than the second order although the difference decreases as the system evolves in time. Overall the decrease of shear pressure with time indicates that the system attempts to return to local equilibration (perfect fluid).

\begin{figure}[h]
\includegraphics[scale=0.31]{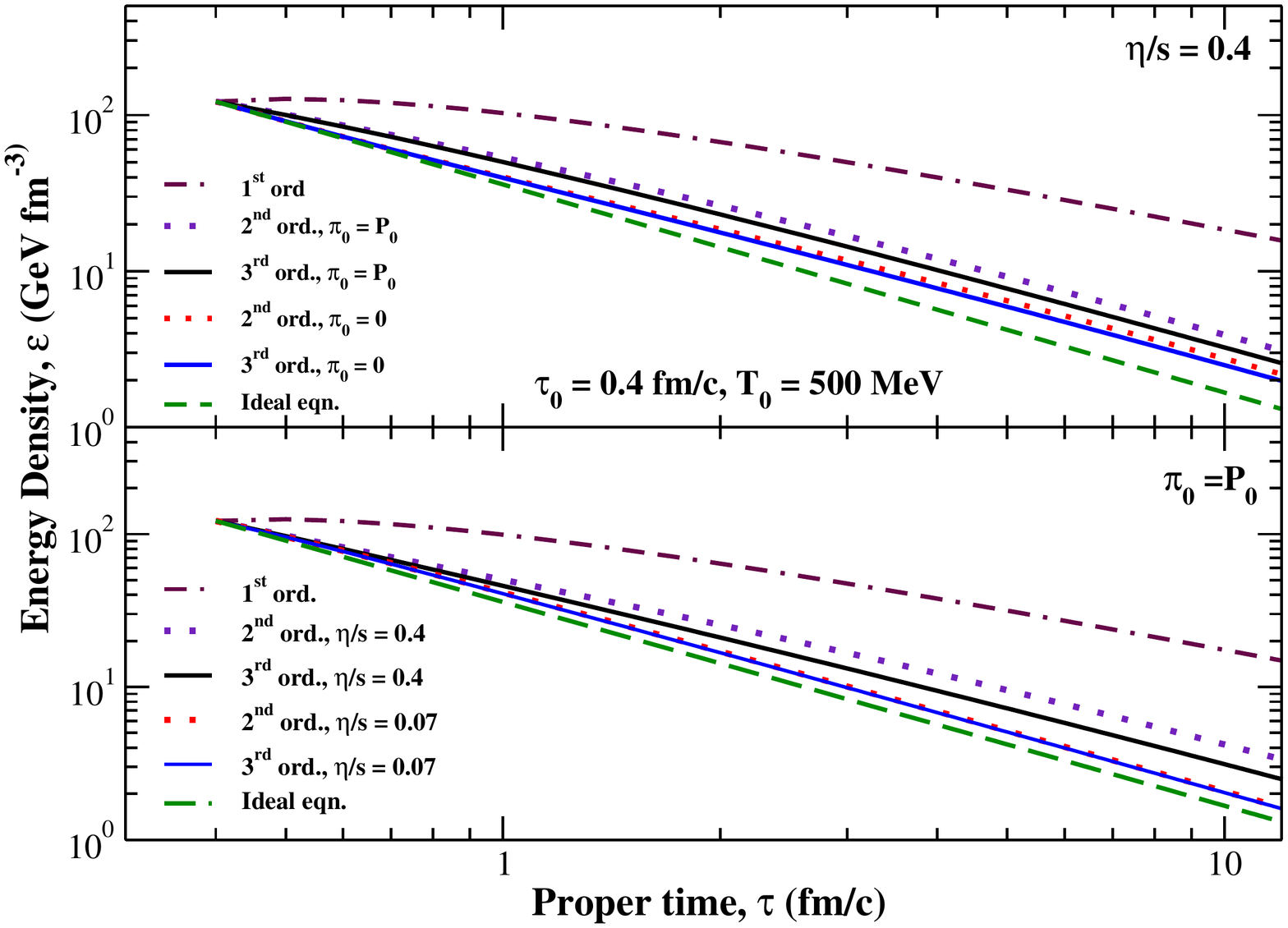}
\caption{\label{figEngDen}(Colour online) Comparison of energy density calculated from second and third order theories with ideal and first order equation. (Top) Two different initial shear values of $\displaystyle{\pi_0}$=0.0 and $\displaystyle{\pi_0}$=$\displaystyle{P_0}$ are used at fixed $\displaystyle{\eta/s}$ = 0.4 (Bottom) Two different $\displaystyle{\eta/s}$ values of 0.07 and 0.4 at fixed $\displaystyle{\pi_0}$ have been used.}
\end{figure}

In Fig.~\ref{figEngDen}, the proper-time evolution of QGP's energy density calculated from ideal, first, second and third order equations has been shown. The top plot shows time evolution of energy density for a moderate value of $\displaystyle{\eta/s}$ = 0.4, and two initial values of $\displaystyle{\pi_0}$ = 0 and P$_0$. The results are shown for an initial temperature  of $T_0$ = 500 MeV and $\tau_0$=0.4 fm/c. The bottom plot keeps $\displaystyle{\pi_0}$ = P$_0$ fixed while varying $\displaystyle{\eta/s}$. The ideal equation in the Bjorken (1+1)D expansion shows the expected trend with energy density decreasing as$\displaystyle{\sim\frac{1}{\tau^{4/3}}}$. The first order equation shows a slight rise in the energy content of the system to a peak until 1 fm/c and then decreases slowly. The peak is more visible in the top plot for the fixed value of $\displaystyle{\eta/s}$. The first order theory differs from ideal scenario by an order of magnitude around 12 fm/c in both plots. The second and third order brings down this difference to values closer to ideal situation. We may also notice from both plots that initial value of $\displaystyle{\pi_0}$ = 0.0 GeV/fm$^3$ or $\displaystyle{\eta/s}$ = 0.07 brings energy density closer to ideal case than $\displaystyle{\pi_0}$ = P$_0$ or $\displaystyle{\eta/s}$ = 0.4.

\begin{figure}[h]
\includegraphics[scale=0.31]{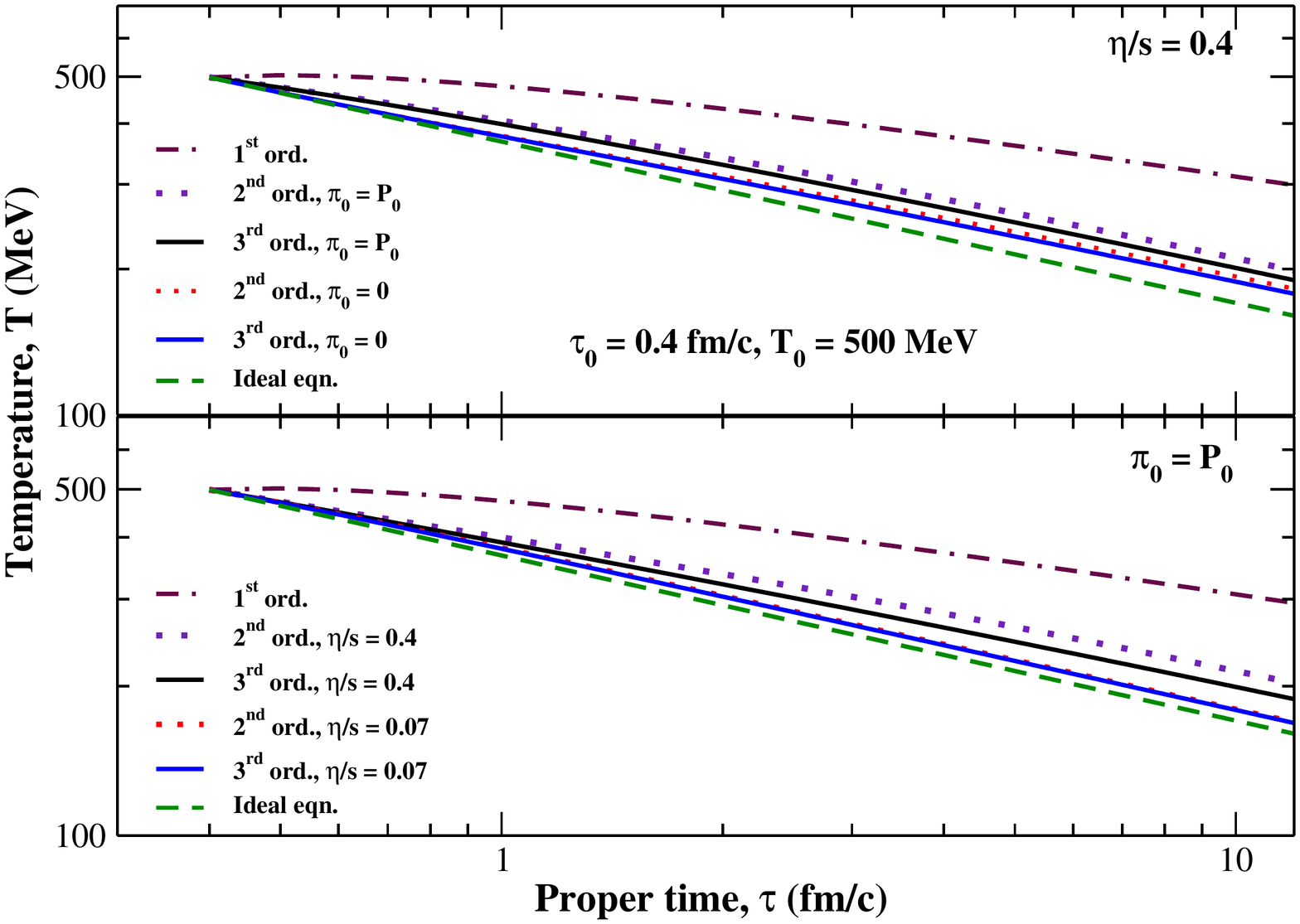}
\caption{\label{figTemp}(Colour online) Comparison of temperature evolution between various orders of dissipative equations at initial LHC temperature of 500 MeV. Two different initial shear values of $\displaystyle{\pi_0}$=0.0 and $\displaystyle{\pi_0}$=$\displaystyle{P_0}$ are used.}
\end{figure}

Similar to Fig.~\ref{figEngDen}, time evolution of temperature of the relativistic fluid is shown in Fig.~\ref{figTemp}. In both plots ideal flow quickly takes initial temperature of 500 MeV to assumed critical temperature of $\displaystyle{T_c}\,\simeq$ 155 - 160 MeV~\cite{Aoki:2009sc,Aoki:2009zzc} (when hadronization sets in) around $\displaystyle{\tau}$=10-12 fm/c approximately. The first order theory is still far above $\displaystyle{T_c}$ by a factor. The ideal and first order theory thus represent two opposite and extreme scenarios of dissipative fluids. In general for the ideal and first order equations, the temperature varies as,
\begin{equation}
\frac{T}{T_0}=\left[\frac{\tau_0}{\tau}\right]^{1/3}\left\lbrace1+\frac{R_0^{-1}}{2}\left(1-\left[\frac{\tau_0}{\tau}\right]^{2/3}\right)\right\rbrace,
\label{temp1}
\end{equation}
where $\displaystyle{T_0\,{\rm and}\,R_0}$ are temperature and Reynolds number at $\displaystyle{\tau\,=\,\tau_0}$. The Reynold number is given by $\displaystyle{\frac{\varepsilon+P}{\pi}}$. With $\displaystyle{R_0^{-1}\,=\,0}$, the temperature evolves in ideal condition (perfect fluid). For ideal fluid, we get $\displaystyle{T\,\simeq\,160}$ MeV at $\displaystyle{\tau}$ = 12 fm/c. Also for $\displaystyle{\tau\,=\,\tau_0}$, the temperature for the two extreme scenarios start from the same point while higher order theories lie in between. 
The third order equations could bring the temperature closest to ideal scenario for initial conditions such as $\,\displaystyle{\pi_0\,=\,0\,,\;\eta/s\,=\,0.4}$ and $\displaystyle{\pi_0\,=\,P_0\,,\;\eta/s\,=\,0.07}$. One may infer that choice of initial conditions  have considerable effects on temperature evolution. Detailed study of transport coefficients used as initial conditions could be carried out using various transport calculations $viz.$ UrQMD, BAMPS, VNI etc. Next we move onto discussion of entropy density.

\begin{figure}[h]
\includegraphics[scale=0.3]{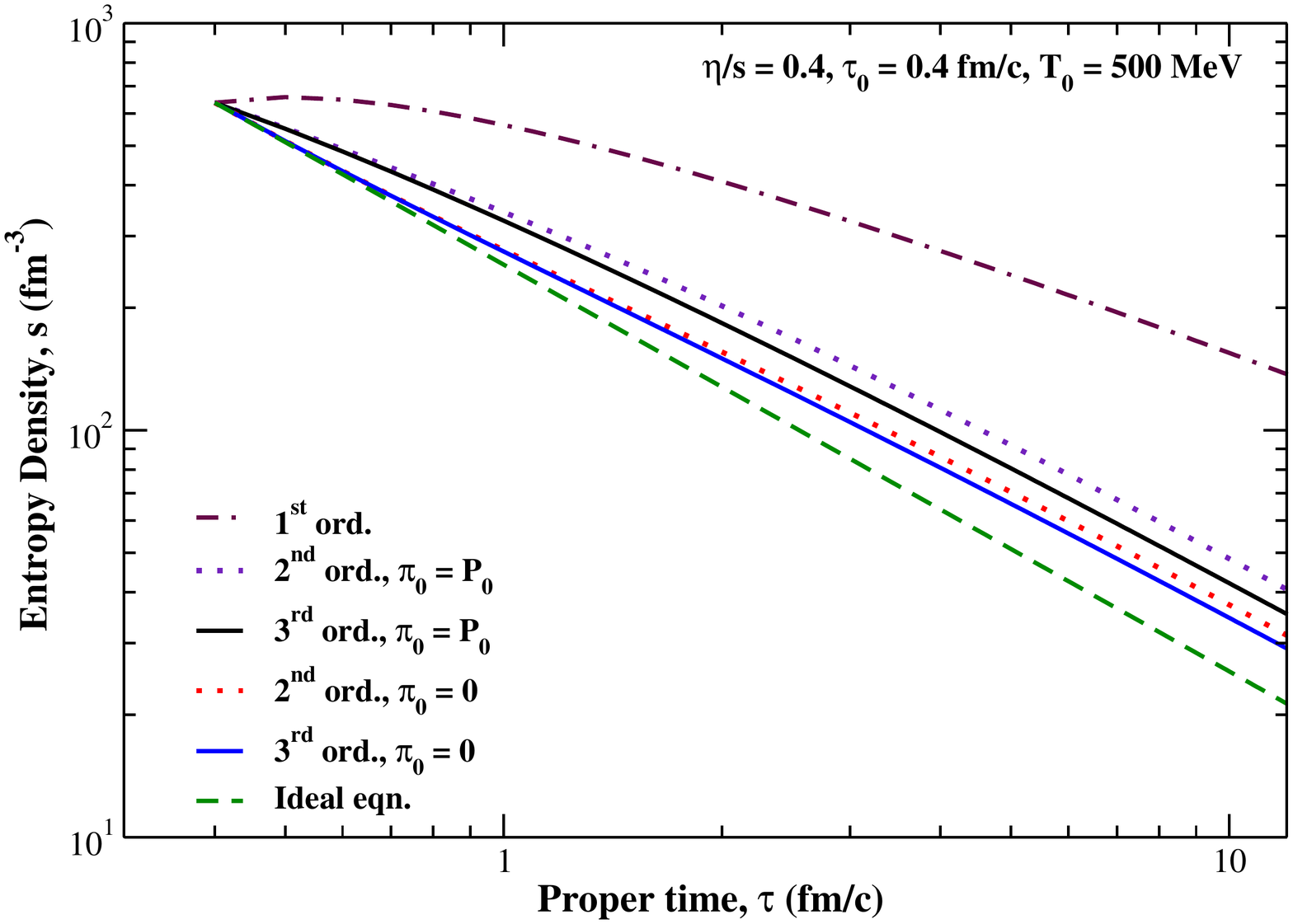}
\caption{\label{figEntDen}(Colour online) proper-time evolution of entropy density for various orders of shear pressure equation.}
\end{figure}

The entropy density, $\displaystyle{s}$ has direct relation to the particle production. The enhancement of production of particles generates more entropy. However, with expansion of the medium or system, as temperature falls off, particle production decreases along with it. Fig.~\ref{figEntDen} shows change in the entropy density with time. Ideal evolution gives a rapid entropy decrease as system nears the critical temperature, assumed to be around 155 - 160 MeV. The first order theories on the other hand decreases most slowly and differs from ideal scenario by an order of magnitude. The third order theory on the other hand brings the system closer to ideal scenario although it differs by a small factor.

\begin{figure}[h]
\includegraphics[scale=0.3]{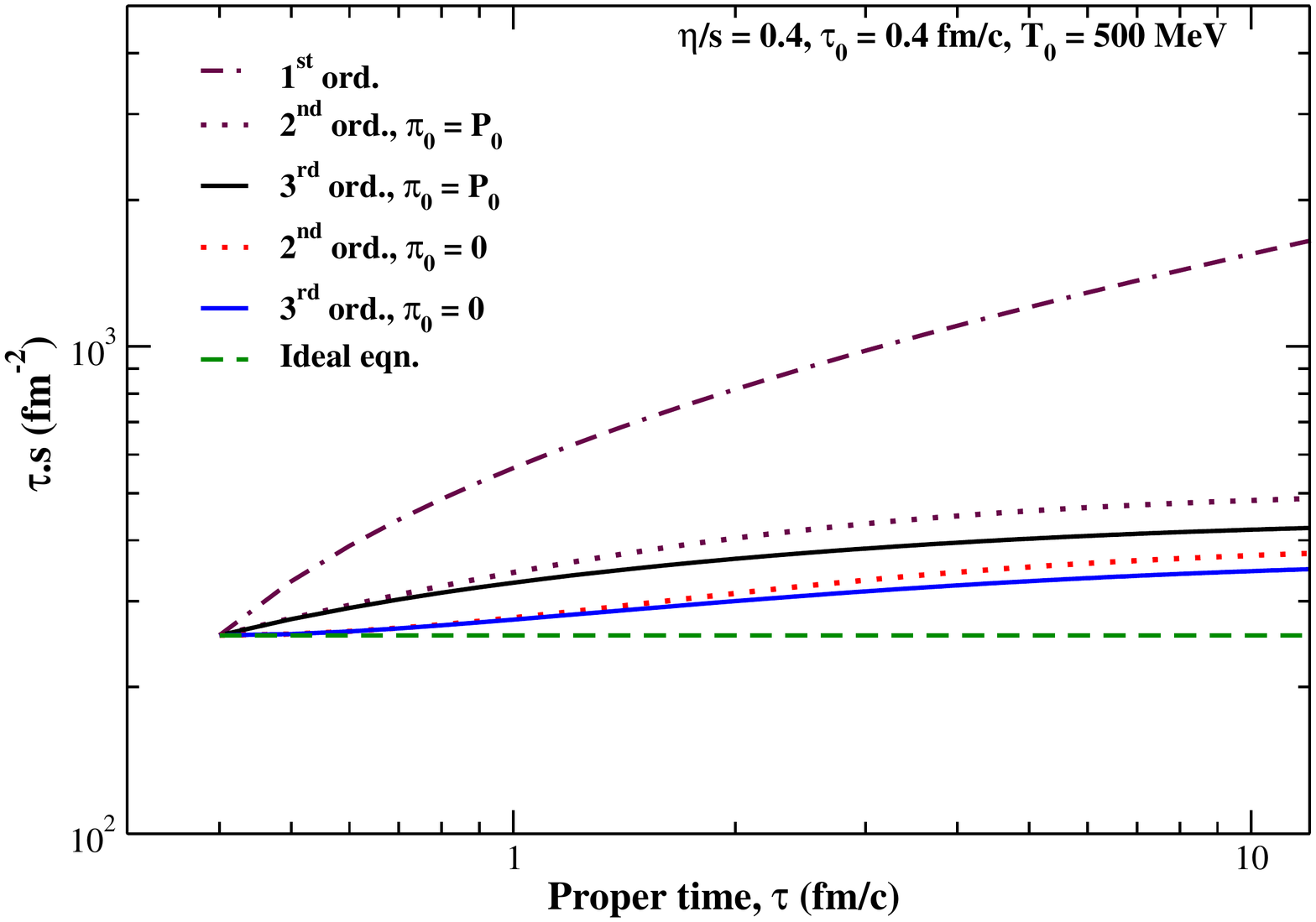}
\caption{\label{figBRC}(Colour online) Time evolution of $\displaystyle{\tau .s}$ with proper time and comparison between orders for $\displaystyle{\pi_0}$ = 0.0 and  $\displaystyle{\pi_0}$ = ${\rm P}_0$.}
\end{figure}

It is known from Bjorken (1+1)D hydrodynamics that in the ideal scenario, $\displaystyle{\tau .s}$ should be conserved over time where $\displaystyle{\tau}$ is the proper time and $\displaystyle{s}$ is the entropy density. The ideal scenario in Fig.~\ref{figBRC} shows the conservation of the quantity as expected from Bjorken's hydrodynamics. The first order theory on the other hand keeps generating more and more entropy which makes the trend rise although there is a small sign of saturation at latter time. The higher orders particularly third order brings this increasing trend down although there is a small increase in the value. However the saturation for the third order develops earlier than the first and second order theories. One may notice that zero initial shear pressure puts the curves closer to ideal situation. The calculation is done for moderate value of $\displaystyle{\eta/s}$ = 0.4 at an initial temperature  of $T_0$ = 500 MeV and $\tau_0$=0.4 fm/c. It can be shown from Bjorken hydrodynamics~\cite{Bjorken:1982qr,Hwa:1985xg,Srivastava:2002ic} that in case of ideal fluid,
\begin{equation}
\tau_0\,.\,s_0=\tau\,.\,s=\frac{2\pi^4}{4\pi\zeta(3)\pi R_T^2}\frac{dN}{dy},
\end{equation}
where $\displaystyle{\frac{dN}{dy}}$ is the observed particle rapidity density distribution and $\displaystyle{R_T}$ is the transverse radius of the system. 
We have assumed chemical freeze-out time to be equal to hadronization or critical time and ignored hadronic medium effects. Although exact analytical expression for 2$\rm{nd}$ and 3$\rm{rd}$ orders is not possible but using Eq.~\ref{temp1} and the expression $\displaystyle{s\,=\,4{\rm a}\,T^3}$ one can have a straightforward expression for ideal and first order,
\begin{eqnarray}
\tau\,.\,s=\tau_0\,.\,s_0\left[1+\left\lbrace\frac{R_0^{-3}}{8}\left(1-\frac{\tau_0^{2/3}}{\tau^{2/3}}\right)^3+\frac{3R_0^{-1}}{2}\right.\right.\nonumber\\\left.\left.\left(1-\frac{\tau_0^{2/3}}{\tau^{2/3}}\right)+\frac{3R_0^{-2}}{4}\left(1-\frac{\tau_0^{2/3}}{\tau^{2/3}}\right)^2\right\rbrace\right],
\end{eqnarray}
where the terms in $\displaystyle{\lbrace ....\rbrace}$ can be termed as first order correction factor to ideal equation. Hence we can write to the first approximation that,
\begin{equation}
\tau\,.\,s=\tau_0\,.\,s_0\left[1+{\rm F}_1(\tau)\right]
\end{equation}
where $\displaystyle{\rm{F_1(\tau)}}$ is the first order correction factor and might be calculated analytically for higher orders. This also shows that correction to multiplicity density can be approximately treated as an additive quantity to ideal scenario.
In our case the calculations have been done for the mid-rapidity where boost-invariance along longitudinal direction is assumed. From the above relation and results from Fig.~\ref{figBRC} at $\displaystyle{\tau}$ = 12 fm/c, one finds for ideal scenario $\displaystyle{\frac{dN_{ch}}{d\eta}}$ = 1749 which is 14\% lower than expt. data $\sim$ 2035 at LHC energy of $\sqrt{s_{NN}}$ = 5.02 TeV~\cite{Adam:2015ptt}. Similarly, for the initial conditions of $\displaystyle{\pi_0}$ = 0 and $\displaystyle{\eta/s}$ = 0.4, third order equation gives $\displaystyle{\frac{dN_{ch}}{d\eta}}$ = 2390 which is about 17\% more than data and 36\% more than the ideal case. Second order equation gives $\displaystyle{\frac{dN_{ch}}{d\eta}}$ = 2578 which is 26\% more than the experimental value and 47\% more than ideal scenario. The first order theory on the other hand gives $\displaystyle{\frac{dN_{ch}}{d\eta}}$ = 11741 which is 7 times higher than ideal value or 500\% increase approx. 

\begin{figure}[h]
\includegraphics[scale=0.31]{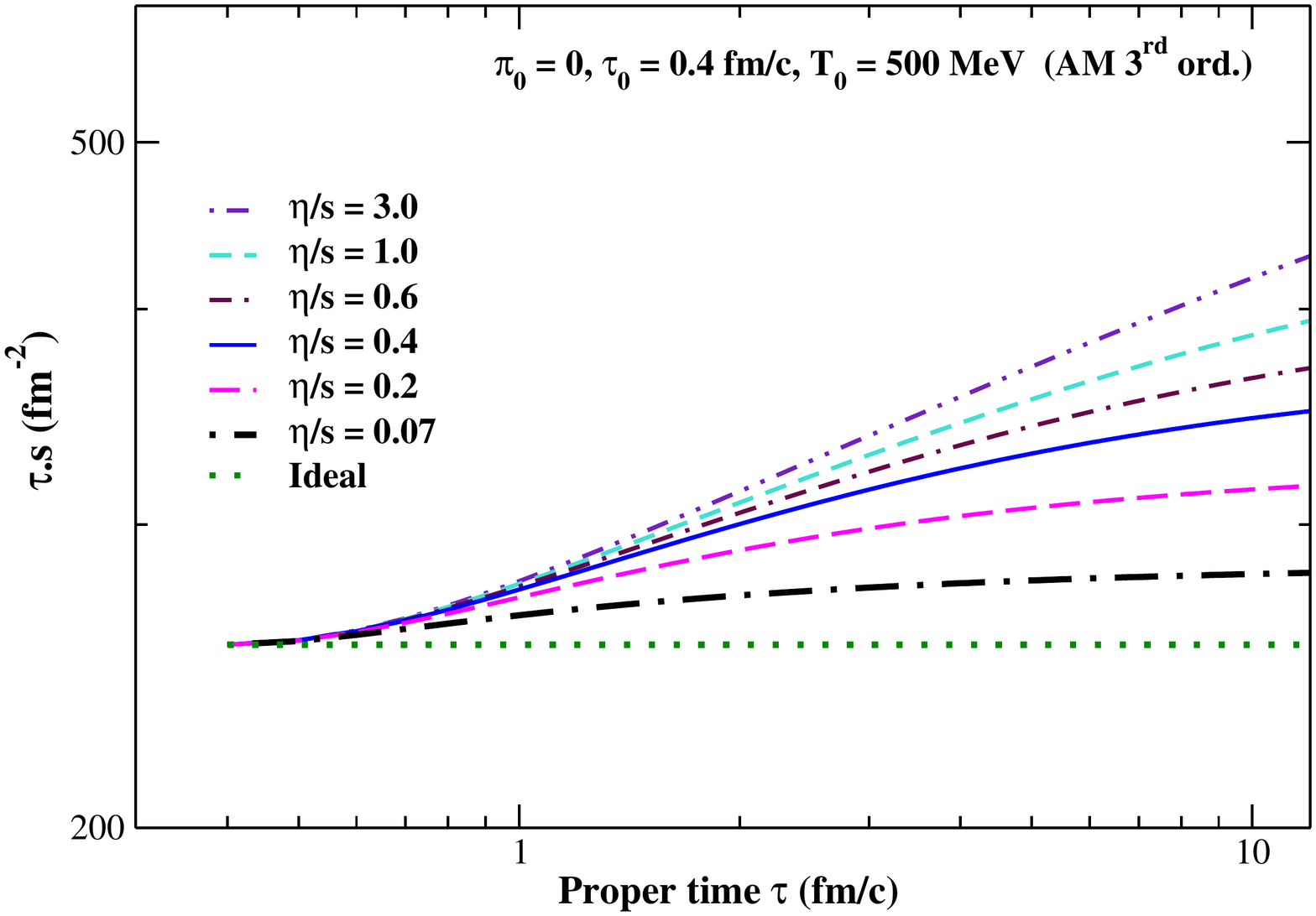}
\caption{\label{figBRetas}(Colour online) Time evolution of $\displaystyle{\tau .s}$ with proper time and comparison between values of $\displaystyle{\eta/s}$ for $\displaystyle{\pi_0}$ = 0.0, $\displaystyle{\tau_0}$ = 0.4 fm/c, and T$_0$ = 500 MeV.}
\end{figure}

As mentioned earlier, initial conditions for transport coefficients play a vital role in precise calculation of entropy or particle production. This is shown in Fig.~\ref{figBRetas} for third order theory. Keeping $\pi_0$ = 0, if $\displaystyle{\eta/s}$ is taken to be 0.07, $\displaystyle{\frac{dN_{ch}}{d\eta}}$ at $\displaystyle{\tau}$ = 12 fm/c is calculated to be 1927 which is 10\% more than ideal eqn. but 5\% less than experimental value. Similarly if we have $\displaystyle{\eta/s}$ = 0.2, we have $\displaystyle{\frac{dN_{ch}}{d\eta}}$ = 2157 which is 23\% more than ideal eqn. and 6\% more than experimental results. Similar calculation with $\displaystyle{\eta/s}$ = 0.07 and 0.2 in second order theory (not shown in figure) gives $\displaystyle{\frac{dN_{ch}}{d\eta}}$ = 1931 and 2222 which are 5\% less  and 10\% more than the expt. value respectively. If one uses first order theory to calculate $\displaystyle{dN_{ch}/d\eta}$, it is found to be 3936 for $\displaystyle{\eta/s}$ = 0.07 at $\displaystyle{\tau}$ = 12 fm/c. It can be seen that first order gives particle production 93\% more than experimental data. Following the above results, one could put limits on the values of $\displaystyle{\eta/s}$ through calculation of particle multiplicity density. However different order theories give a slight different range for the parameter although part of these ranges overlap. On the other hand, such dissimilarities might also limit the universal adaptation of a particular ordered theory for dissipative fluids. This aspect would be further investigated and reported. 

It could be recalled that no effects due to dissipative heat flow as well as from bulk viscous pressure have been included in the present calculations. Bjorken scaling solution excludes the heat flow coefficient and ultra-relativistic scenario neglects effects of bulk viscosity. Consequently most of the terms in Eq.~\ref{entropy8} are absent and gives us a very simplistic picture. The goal of developing the third order viscous hydro-equations is to highlght the coupling coefficients $\displaystyle{S_n^m}$s which display not only the correlation of dissipative fluxes among themselves but also with each other. Such correlations are absent in second order theories. In our current work Bjorken scenario has eliminated the possibilities of correlations among fluxes except shear viscous pressure term, $\displaystyle{\pi_{\alpha\beta}\pi^{\alpha\beta}}$. The  Inclusions of the other two important fluxes and all type of correlations should bring in changes in the production of entropy (see Eqn.~\ref{entropyrel1}), energy density, relaxation time and other rich components and these effects must be extensively studied in near future. No hadronic medium effects have been included in our simple model. It naively assumes chemical freeze-out scenario occurs at the critical temperature itself. Relaxing this condition should bring in extra particle production from hadronic medium mainly from hadron decay channels. Earlier studies on temperature evolution on pion gas with second order theory has already been done in ~\cite{Muronga:2003ta}.

\section{Conclusions}

In the present manuscript third order shear relaxation equation in Bjorken scenario has been developed following earlier second order calculation by A. Muronga. Comparison to other third order models in similar scaling solutions by A. El et al. and A. Jaiswal have been done. Consequently we have worked on to check consistency of our calculations in a very simplistic scenario. This would serve as a test bench for future development into (3+1)D dissipative hydrodynamics. The differences in the results have been highlighted and discussed. The coupling coefficients in the relaxation equations from these models have been found to be slightly different from our calculations and the output is found to be sensitive to values of the coefficients. The difference also indicates one of the possible source of uncertainties in the output and thus must be precisely evaluated. More detailed investigation on the effects of the coefficients on thermodynamic variables are being carried out. Comparison with transport theory of BAMPS has been carried out via $\displaystyle{P_L/P_T}$ ratio plot. The BAMPS data has similar trend to the results from AM model. However the present method of using Grad's 14 moment approximation shows considerable difference with transport results especially at high $\displaystyle{\eta/s}$ values. This would be investigated and reported in future. Non-linear terms such as $\displaystyle{\partial_i S_3^2\,\rm{and}\,\partial_i S_5^3}$ associated with shear viscous pressure have been included in the relaxation equation and have shown effects on the calculated thermodynamic quantities at low $\displaystyle{\tau}$ or for higher $\displaystyle{\eta/s}$ values. However because of the absence of coupling coeffcients associated with other fluxes or correlated terms, the final outcome on the observed thermodynamic quantities is not conclusive. The assumptions are being relaxed and the precise determination of the coupling coefficients $\displaystyle{S^m_n}$ and effect of EoS on them are being currently studied and will be reported soon. Third order correction to shear relaxation time $\displaystyle{\tau_\pi}$ has explicit effects of shear pressure $\displaystyle{\pi}$ which was absent in the second order. Also third order relaxation time $\displaystyle{\tau_\pi^{(3)}}$ is more than $\displaystyle{\tau_\pi}$ but appears to converge to it at latter time. It would be interesting to study what effects ordered theories higher than 3 would bring on the relaxation time. 

The ratio of longitudinal pressure to transverse pressure ratio calculated for $\displaystyle{\eta/s}\,>$ 0.5 shows an almost flat and horizontal trend with time. This might indicate that system with high viscosity may not go back to equilibrium during QGP lifetime. Consequently a detailed study of time-scale or system length-scale dependence of relaxation time of dissipative fluxes and correlation between the thermodynamic variables and transport coefficients could be carried out. This would also bring out in-depth information on ordere theories and correlation between system evolution, hadronization and freeze-out times and various coupling coefficients. A precise determination of transport coefficients such as shear $\displaystyle{\eta}$ and bulk $\displaystyle{\zeta}$ viscosities and thermal conductivity $\displaystyle{\kappa}$ must also be carried out. Parton cascade and transport models like UrQMD, BAMPS, VNI/BMS etc. could help us in the study. The study of time evolution of thermodynamic variables and transport coeffcients would also help us develop the equation of state (EoS) which is vital for transport simulation~\cite{Zhang:1998tj,Pal:2000zm,Bass:2002fh,Muronga:2007ny,De:2010zg,Fuini:2010xz,Muronga:2003tb,Xu:2004mz,Xu:2007ns,Wesp:2011yy,Chen:2018mwr}. 

It is also seen from first order theory, that there is a large increase in temperature and entropy production. They are almost an order in magnitude different from ideal equation results. Both second and third order theories bring down this difference. As mentioned in~\cite{El:2009vj}, it would be also an interesting study to go beyond third order in present calculations to study the oscillatory nature of ordered theories on thermodynamic variables. Calculations needed to be done with full transport theories which include dissipative heat flow and bulk viscous pressure along with shear pressure and extend our theories to full dissipative hydrodynamics. One needs to study the changes in variables due to the additional effects of heat flow and bulk viscous pressure. Inclusion of bulk viscosity becomes important for the heavy ion collision systems with massive particles, large chemical potential or other small systems from proton-proton collisions. Also our current study has been carried out for a single initial temperature of 500 MeV. Other values for initial temperature could be included to further elucidate temperature dependences of dissipative fluxes, entropy production etc. Hadron medium effects weren't included in this simple model. Works are being carried out although some results for hadronic regime using second order theories have already been shown by A. Muronga et al.~\cite{Muronga:2003ta}. One of the many purposes of determining hydrodynamic attractors is to test the convergence of hydrodynamics coefficients to an arbitrary high order. It shows the viability and applicability of present form of relativistic fluid mechanics we are using today. The techniques could be extended to both massless and massive particles. A part of the current work could also be directed along the study of attractors in near future.

\begin{acknowledgments}
One of the authors (MY) wishes to acknowledge the support of Department of Physics, Nelson Mandela University in allowing us its facilities for the work. The author (MY) is grateful to Research Capacity Development, Nelson Mandela University for supporting the project.
The authors would like to thank Prof. Tomoi Koide for his valuable suggestions, and Dr. Amaresh Jaiswal for his valuable discussions on methods and providing us with BAMPS data from A. El et al's paper.
\end{acknowledgments}

\appendix
\section{}
Some of the mathematical terms used in our calculations and shown in the manuscript have been explained here briefly. The deviation function $\displaystyle{\phi(x)}$ in Eq.~\ref{distribution} is written as function of moments $\displaystyle{\epsilon,\,\epsilon_\mu\,\rm{and}\,\epsilon_{\mu\nu}}$,
\begin{eqnarray}
\phi(x)&=&\epsilon-\epsilon_\mu p^\mu+\epsilon_{\mu\nu}p^\mu p^\nu
\end{eqnarray}
where the moments are function of dissipative fluxes as follows
\begin{eqnarray}
\epsilon&=&A_0\nonumber\\
\nonumber\\
\epsilon_\mu&=&A_1u_\mu\Pi - B_0 q_\mu \nonumber\\
\nonumber\\
\epsilon_{\mu\nu}&=&A_2(4u_\mu u_\nu -g_{\mu\nu})\Pi -B_1u_\mu q_\nu + C_0\pi_{\mu\nu}\nonumber\\
\end{eqnarray}

The coeffcients in the moments in $\phi(x)$ are calculated to be,
\begin{eqnarray}
A_0&=&3A_2D^{-1}_{20}(D_{30}+I_{41}I_{20}-I_{30}I_{31})\nonumber\\
\nonumber\\
A_1&=&3A_2D^{-1}_{20}(4(I_{10}I_{41}-I_{20}I_{31}))\nonumber\\
\nonumber\\
A_2&=&\frac{1}{4I_{42}\Omega}\nonumber\\
\nonumber\\
B_0&=&B_1\frac{I_{41}}{I_{31}}\nonumber\\
\nonumber\\
B_1&=&\frac{1}{\Lambda I_{21}}\nonumber\\
\nonumber\\
C_0&=&\frac{1}{2I_{42}}
\end{eqnarray}

with
\begin{equation}
\Lambda=\frac{D_{31}}{I^2_{21}}
\end{equation}
and
\begin{eqnarray}
\Omega =&-&\frac{I_{10}}{D_{20}I_{31}}\left[I_{30}\left(I_{30}-\frac{I_{31}}{I_{21}I_{20}}\right)-I_{40}\left(I_{20}-\frac{I_{31}}{I_{21}}I_{10}\right)\right]\nonumber\\
&+&\beta\frac{I_{41}}{I_{31}}
\end{eqnarray}
The integration $\displaystyle{I_{nk}}$ is a scalar coefficient that depends on the equilibrium distribution parameters $\displaystyle{\alpha\,\rm{and}\,\beta}$. It appears in the n$^{\rm{th}}$ moment of the distribution function as (see Ref.~\cite{Muronga:2006zx})
\begin{eqnarray}
I^{\mu_1\,.....\mu_n}(x)&=&\int dw\,p^\mu_1....p^\mu_n\,f(x,p)\nonumber\\
                        &=&\sum_{k=0}^i\,a_{nk}\,I_{nk}\Delta^{(2k}u^{n-2k)}
\end{eqnarray}
where $\displaystyle{I_{nk}}$ is given by
\begin{equation}
I_{nk}=\frac{A_0}{(2k+1)!!}\int{dw(p^\alpha u_\alpha-p^\alpha p_\alpha)^k(p^\mu u_\mu)}^{n-2k}\frac{1}{\text{e}^{\beta_\mu p^\mu-\alpha}-a}\\
\end{equation}

The quantity, $\displaystyle{D_{nk}}$ can be shown to be
\begin{equation}
D_{nk}=I_{n+1,k}I_{n-1,k}-I^2_{n,k}
\end{equation}

In ultra-relativistic limit, as rest mass, $m\longrightarrow\,0$ w.r.t. particles' kinetic energy, it can be shown 

\begin{equation}
I_{nk}=\frac{4\pi A_0}{(2k+1)!!}T^{n+2}\int\limits^\infty_0{dx\,x^{n+1}\frac{1}{e^{x-\phi}}}\,,
\end{equation}
with $\displaystyle{x=\frac{p}{T}}$ and $\displaystyle{z=\frac{m}{T}}$ are the new variables. $\displaystyle{\phi}$ being the chemical potential has been taken to be zero here.
Then we have,
\begin{equation}
I_{nk}=\frac{4\pi A_0}{(2k+1)!!}T^{n+2}(n+1)!.\,.
\end{equation}

The quantities $\displaystyle{\Omega}$ that come in the bulk viscosity and $\displaystyle{\Lambda}$ that appears in the thermal conductivity coefficient can now be reduced to,
\begin{equation}
\Omega=0\,,\;\; \Lambda=4T^2.
\end{equation}

Using Eqs.~\ref{entropy5}--\ref{entropy7}, the coefficients for the second order in the entropy 4-current, Eq.\ref{entropy8} have been calculated as follows,
\begin{eqnarray}
S^2_1&=&\frac{3}{\beta I^2_{42}\Omega^2}(5I_{52}-\frac{3}{D_{20}}[I_{31}(I_{31}I_{30}-I_{41}I_{20})\nonumber\\
&+& I_{41}(I_{41}I_{10}-I_{31}I_{20})])\nonumber\\
\nonumber\\
S^2_2&=&\frac{D_{41}}{\beta\Lambda I^2_{21}I_{31}}\nonumber\\
\nonumber\\
S^2_3&=&\frac{1}{2}\frac{I_{52}}{\beta I^2_{42}}\nonumber\\
\nonumber\\
S^2_4&=&\frac{D_{41}D_{20}-D_{31}D_{30}}{\beta\Lambda\Omega I_{42}I_{21}I_{31}D_{20}}\nonumber\\
\nonumber\\
S^2_5&=&\frac{I_{31}I_{52}-I_{41}I_{42}}{\beta\Lambda I_{42}I_{21}I_{31}}
\end{eqnarray}

The coefficients for the third order dissipative quantities in entropy 4-current, Eq.\ref{entropy8} have similarly been calculated as followa,
\begin{eqnarray}
S^3_1&=&\frac{1}{\beta}[I_{10}A_0^3-3I_{20}A_0^2A_1+9I_{30}A_0^2A_2-I_{40}A_1(A_1^2+18A_0A_2)\nonumber\\&+&9I_{50}A_2(3A_0A_2-A_1^2)-27I_{60}A_1A_2^2+27I_{70}A_2^3]\nonumber\\
\nonumber\\
S^3_2&=&-\frac{1}{\beta}[3I_{41}B_0(2A_0B_1+A_1B_0)+3I_{51}(2A_1B_0B_1-A_0B_1^2\nonumber\\
&+&3A_2B_0^2)+3I_{61}(6A_2B_0B_1+A_1B_1^2)-9I_{71}A_2B_1^2]\nonumber\\
\nonumber\\
S^3_3&=&\frac{1}{\beta}[6I_{52}A_0C_0^2-6I_{62}A_1C_0^2+18I_{72}A_2C_0^2]\nonumber\\
\nonumber\\
S^3_4&=&\frac{3C_0}{\beta}[I_{52}B_0^2-2I_{62}B_0B_1+I_{72}B_1^2]\nonumber\\
\nonumber\\
S^3_5&=&\frac{2I_{73}C_0^3}{\beta}\nonumber\\
\nonumber\\
S^3_6&=&\frac{3}{\beta}[I_{21}A_0^2B_0+I_{31}A_0^2B_1+I_{41}(A_1^2B_0+2A_0A_1B_1\nonumber\\
&+&6A_0A_2B_1)+I_{51}(A_1^2B_1-6A_0A_2B_1+6A_1A_2B_0)\nonumber\\
&+&3I_{61}A_2(2A_1B_1+3A_2B_0)-9I_{71}A_2^2B_1]\nonumber\\
\nonumber\\
S^3_7&=&-\frac{1}{\beta}[I_{42}B_0^3-3I_{52}B_0^2B_1+3I_{62}B_0B_1^2-I_{72}B_1^3]\nonumber\\
\nonumber\\
S^3_8&=&\frac{3}{\beta}[I_{73}B_1C_0^2-I_{63}B_0C_0^2]\nonumber\\
\nonumber\\
S^3_9&=&-\frac{1}{\beta}[I_{42}A_0B_0C_0-I_{52}A_1B_0C_0+I_{52}A_0B_1C_0\nonumber\\
&+&I_{62}A_1B_1C_0+3I_{62}A_2B_0C_0-3I_{72}A_2B_1C_0]\nonumber\\
\nonumber\\
S^3_{10}&=&\frac{1}{\beta}[I_{73}B_1C_0^2-I_{63}B_0C_0^2]\nonumber\\
\end{eqnarray}

In the ultra-relativistic limits, the coeffcients can be shown to be, for eg.
\begin{eqnarray}
\Lambda &\rightarrow & \frac{1}{2}P^{-1}\,,\;S^2_2\rightarrow\frac{5}{4}P^{-1}\,,\;S^2_3\rightarrow\frac{3}{4}P^{-1}\,,\;S^2_5\rightarrow\frac{1}{7}P^{-1}\,,\nonumber\\
S^3_4 & \rightarrow & 6P^{-2}\,,\;S^3_5\rightarrow\frac{3}{4}P^{-2}\,,\;S^3_7\rightarrow 2P^{-2}\,,\;S^3_{10}\rightarrow\frac{9}{32}P^{-2}\,\, {\rm etc.}\nonumber\\
& \rm{where} & \,,P=I_{21}=\frac{4\pi A_0}{3}T^4 3! \;\; \rm{is\; the\; pressure\,.}
\end{eqnarray}
\vskip 1.2in
\section{}
Using the entropy principle $\displaystyle{\partial_\mu S^\mu\geq\,0}$ and Eq.\ref{entropy8} the third order expression for the dissipative fluxes been calculated as
\begin{eqnarray}
\Pi = &-&\zeta\left[\nabla_\alpha u^\alpha + 2S_1^2\dot{\Pi}+S_4^2\nabla_\alpha q^\alpha+\Pi(\dot{S_1^2}+S_1^2\nabla_\alpha u^\alpha)\right.\nonumber\\
&+&\left. q^\alpha(\nabla_\alpha S_4^2-S_4^2\dot{u^\alpha})+3S_1^3\dot{\Pi}\,\Pi +2S_2^3\dot{q_\alpha}q^\alpha\right.\nonumber\\
&+&\left. 2S_3^3\dot{\pi}_{\langle\alpha\beta\rangle}\pi^{\langle\alpha\beta\rangle}
+S_6^3(\Pi\nabla_\alpha q^\alpha+ q^\alpha\nabla+\alpha\Pi)\right.\nonumber\\
&+&\left. S_9^3(\pi^{\langle\alpha\beta\rangle}\nabla_\alpha q_\beta +q_\beta\nabla_\alpha\pi^{\langle\alpha\beta\rangle})\right.\nonumber\\
&+&\left. \Pi^2(\dot{S_1^3}+S_1^3\nabla_\alpha u^\alpha)-q^\alpha q_\alpha(\dot{S_2^3}+S_2^3\nabla_\alpha u^\alpha)\right.\nonumber\\
&+&\left. \pi^{2\langle\alpha\beta\rangle}(\dot{S_3^3}+S_3^3\nabla_\alpha u^\alpha)+\Pi q^\alpha(\nabla_\alpha S_6^3-S_6^3\dot{u_\alpha})\right.\nonumber\\
&+&\left. \pi^{\langle\alpha\beta\rangle}q_\beta(\nabla_\alpha S_9^3-S_9^3\dot{u_\alpha})\right]\nonumber\\
\end{eqnarray}

\begin{eqnarray}
q^\alpha &=& \kappa T\Delta^{\alpha\mu}\left[\left(\frac{\nabla_\alpha T}{T}-\dot{u_\alpha}\right)+2S_2^2\dot{q}_\alpha + S_4^2\nabla_\alpha\Pi\right.\nonumber\\
&+&\left. S_5^2\nabla^\beta\pi_{\langle\alpha\beta\rangle}+q_\alpha(\dot{S_2^2}+S_2^2\nabla_\nu u^\nu)+\Pi(\nabla_\alpha S_4^2-S_4^2\dot{u_\alpha})\right.\nonumber\\
&+&\left. \pi_{\langle\alpha\beta\rangle}(\nabla^\beta S_5^2-S_5^2\dot{u^\beta})-S_2^3(2\Pi\dot{q_\alpha}+q_\alpha\dot{\Pi})\right.\nonumber\\
&+&\left. S_4^3(2\dot{q^\beta}\pi_{\langle\alpha\beta\rangle}+q^\beta\dot{\pi}_{\langle\alpha\beta\rangle})+2S_6^3\Pi\nabla_\alpha\Pi\right.\nonumber\\
&-&\left. 2S_7^3q^\beta\nabla_{q_\beta}+S_9^3(\Pi\nabla^\beta\pi_{\langle\alpha\beta\rangle}+\pi_{\langle\alpha\beta\rangle}\nabla^\beta\Pi)\right.\nonumber\\
&+&\left. 2S_{10}^3\pi_{\langle\beta\nu\rangle}\nabla_\alpha\pi^{\langle\alpha\nu\rangle}-\Pi q_\alpha(\dot{S_2^3}+S_2^3\nabla_\nu u^\nu)\right.\nonumber\\
&+&\left. q^\beta\pi_{\langle\alpha\beta\rangle}(\dot{S_4^3+S_4^3\nabla_\nu u^\nu})+\Pi^2\nabla_\alpha S_6^3\right.\nonumber\\
&-&\left. q^\lambda q_\lambda\nabla_\alpha S_7^3+\pi^{2\langle\lambda\lambda\rangle}\nabla_\alpha S_8^3+2S_8^3\pi^{\langle\mu\nu\rangle}\nabla_\alpha\pi_{\langle\mu\nu\rangle}\right.\nonumber\\
&+&\left. \Pi\pi_{\langle\alpha\beta\rangle}(\nabla^\beta S_9^3-S_9^3\dot{u^\beta})+\pi_{\langle\alpha\beta\rangle}^2(\nabla^\beta S_{10}^3-S_{10}^3\dot{u^\beta})\right.\nonumber\\
&+&\left. S_7^2q_\alpha q^\lambda q_\lambda\right]\nonumber\\
\end{eqnarray}

\begin{eqnarray}
\pi^{\langle\mu\nu\rangle}&=&2\eta\Delta^{\alpha\mu}\Delta^{\beta\nu}\left[\nabla_{\langle\alpha}u_{\beta\rangle}+2S_3^2\dot{\pi_{\langle\alpha\beta\rangle}}\right.\nonumber\\&+&\left. S_5^2\nabla_{\langle\alpha}q_{\beta\rangle}+\pi_{\langle\alpha\beta\rangle}(\dot{S_3^2}+S_3^2\nabla_\lambda u^\lambda)\right.\nonumber\\
&+&\left. q_{\langle\alpha}(\nabla_{\beta\rangle}S_5^2-S_5^2\dot{u_{\beta\rangle}})+S_3^3(2\Pi\dot{\pi}_{\langle\alpha\beta\rangle}+\pi_{\langle\alpha\beta\rangle}\dot{\Pi})\right.\nonumber\\
&+&\left. 2S_4^3\dot{q}_{\langle\alpha}q_{\beta\rangle}+3S_5^3\dot{\pi}_{\langle\alpha\lambda\rangle}\pi^{\langle\lambda}_{\beta\rangle}+S_8^3\pi_{\langle\alpha\beta\rangle}\nabla_\lambda u^\lambda\right.\nonumber\\
&+&\left. S_9^3(\Pi\nabla_{\langle\alpha}q_{\beta\rangle}+q_{\langle\alpha}\nabla_{\beta\rangle}\Pi)+S_{10}^3(q_{(\alpha}\nabla^\lambda\pi_{\langle\alpha)\lambda\rangle}\right.\nonumber\\
&+&\left. \pi_{\langle\lambda(\alpha\rangle}\nabla^\lambda q_\beta)+\Pi\pi_{\langle\alpha\beta\rangle}(\dot{S_3^3}+S_3^3\nabla_\lambda u^\lambda)\right.\nonumber\\
&+&\left. q_{\langle\alpha}q_{\beta\rangle}(\dot{S_4^3}+S_4^3)\nabla_\lambda u^\lambda+\pi_{\langle\alpha\lambda\rangle}\pi^{\langle\lambda}_{\beta\rangle}(\dot{S_5^3}+S_5^3\nabla_\lambda u^\lambda)\right.\nonumber\\
&+&\left. \pi_{\langle\alpha\beta\rangle}q^\lambda(\nabla_\lambda S_8^3-S_8^3\dot{u}_\lambda)+\Pi q_{\langle\alpha}(\nabla_{\beta\rangle}S_9^3-S_9^3\dot{u}_{\beta\rangle})\right.\nonumber\\
&+&\left. \pi_{\langle\alpha\lambda}q^\lambda(\nabla_{\beta\rangle}S_{10}^3-S_{10}^3\dot{u}_{\beta\rangle})\right]\nonumber\\
\end{eqnarray}
 Thus upto third order the bulk, heat and shear equations are the sum of the zeroth, first, second and third order contributions.


\end{document}